\documentclass[twoside,10pt]{article}

\usepackage{multicol}
\usepackage{amsmath}
\usepackage{amsfonts} 
\usepackage{eucal}
\usepackage{cite}
\usepackage{fancyhdr}
\renewcommand{\sectionmark}[1]%
 {\markboth{\thesection:\ #1}{}}

\numberwithin{equation}{section}

\begin{document}

%\draft

\title{\bfseries Mathematical Properties of a \\ 
New Levin-Type Sequence Transformation \\
Introduced by \v{C}\'{\i}\v{z}ek, Zamastil, and Sk\'{a}la. \\
I. Algebraic Theory}

\pagenumbering{roman}

\author{Ernst Joachim Weniger \\
Institut f\"ur Physikalische und Theoretische Chemie \\
Universit\"at Regensburg, D-93040 Regensburg, Germany \\
joachim.weniger@chemie.uni-regensburg.de}

\date{Submitted to Journal of Mathematical Physics \\
Date of Submision: 25 June 2003}

\maketitle

\bigskip

\centerline{\textbf{PACS numbers:} 02.30.Lt, 02.30.Mv, 02.60.-x}

\bigskip

\begin{abstract}
\noindent
\v{C}\'{\i}\v{z}ek, Zamastil, and Sk\'{a}la [J.\ Math.\ Phys.\
\textbf{44}, 962 -- 968 (2003)] introduced in connection with the 
summation of the divergent perturbation expansion of the hydrogen atom
in an external magnetic field a new sequence transformation which uses
as input data not only the elements of a sequence $\{ s_n
\}_{n=0}^{\infty}$ of partial sums, but also explicit estimates $\{
\omega_n \}_{n=0}^{\infty}$ for the truncation errors. The explicit
incorporation of the information contained in the truncation error
estimates makes this and related transformations potentially much more
powerful than for instance Pad\'{e} approximants. Special cases of the
new transformation are sequence transformations introduced by Levin
[Int.\ J.\ Comput.\ Math.\ B \textbf{3}, 371 -- 388 (1973)] and Weniger
[Comput.\ Phys.\ Rep.\ \textbf{10}, 189 -- 371 (1989), Sections 7 -9;
Numer.\ Algor.\ \textbf{3}, 477 -- 486 (1992)] and also a variant of
Richardson extrapolation [Phil.\ Trans.\ Roy.\ Soc.\ London A
\textbf{226}, 299 -- 349 (1927)]. The algebraic theory of these 
transformations -- explicit expressions, recurrence formulas, explicit
expressions in the case of special remainder estimates, and asymptotic
order estimates satisfied by rational approximants to power series -- is
formulated in terms of hitherto unknown mathematical properties of the
new transformation introduced by \v{C}\'{\i}\v{z}ek, Zamastil, and
Sk\'{a}la. This leads to a considerable formal simplification and
unification.
\end{abstract}

\newpage 
\tableofcontents

\newpage 
\pagenumbering{arabic} 

\typeout{==> Introduction}
\section{Introduction}
\label{Sec:Intro}

The most important and most versatile systematic approximation method in
quantum physics is eigenvalue perturbation theory (see for example
\cite{Simon/1991}). Thus, the question, whether perturbation expansions 
converge or diverge, is of principal importance. Already in 1952, Dyson
\cite{Dyson/1952} had argued that perturbation expansions in quantum
electrodynamics should diverge. Around 1970, Bender and Wu
\cite{Bender/Wu/1969,Bender/Wu/1971,Bender/Wu/1973} showed in their work
on anharmonic oscillators that factorially divergent perturbation
expansions occur also in nonrelativistic quantum mechanics. In the
following years, many other quantum systems were investigated, and in
the overwhelming majority factorially divergent perturbation expansions
were found (see for example \cite[Table 1]{Fischer/1997} or the articles
reprinted in \cite{LeGuillou/Zinn-Justin/1990}). Consequently, summation
methods are needed to give the divergent perturbation series of quantum
physics any meaning beyond mere formal expansions and to extract
numerical information from them. A very readable discussion of the
usefulness of summation and related techniques from a physicist's point
of view can be found in the monograph by Bender and Orszag
\cite{Bender/Orszag/1978}.

Factorially divergent power series occur also in asymptotic expansions
for special functions. However, special functions can normally be
computed via a variety of different representations. Accordingly, in
mathematics there is usually no compelling need to use divergent series
for computational purposes, whereas in quantum physics it is frequently
quite difficult or even impossible to find alternatives to divergent
perturbation expansions. Consequently, summation techniques are far more
important in physics than in mathematics. Nevertheless, the evaluation
of special functions by summing divergent asymptotic expansions can be
remarkably effective (see for example
\cite{Weniger/1990,Weniger/1994b,Weniger/1996d,Weniger/Cizek/1990}).

In physics, the best known and most widely used summation techniques are
Borel summation \cite{Borel/1899,Borel/1928}, which replaces a divergent
perturbation expansion by a Laplace-type integral, and the method of
Pad\'{e} approximants \cite{Pade/1892}, which transforms the partial
sums of a power series to a rational function. Both approaches have been
remarkably successful, but they have -- like all other numerical
techniques -- certain shortcomings and limitations. For example, the
Borel method is very powerful, but conceptually and computationally very
demanding. From a technical point of view, Pad\'{e} approximants can be
applied remarkably easily, but they are not necessarily powerful enough
to sum all perturbation series of interest. For example, Graffi and
Grecchi \cite{Graffi/Grecchi/1978} showed rigorously that Pad\'{e}
approximants are not able to sum the perturbation expansion of the octic
anharmonic oscillator whose series coefficients grow roughly like
$(3n)!/n^{1/2}$ \cite[Eq.\ (3)]{Bender/Wu/1971}. Accordingly, it is
worth while to look for alternative techniques which are at least in
some cases capable of producing better summation results.

Pad\'{e} approximants accomplish an acceleration of convergence or a
summation by converting the partial sums of a power series to a doubly
indexed sequence of rational functions. This is also done by other,
albeit less well known nonlinear transformations (see for example
\cite{Brezinski/1977,Brezinski/1978,Brezinski/RedivoZaglia/1991a,%
Wimp/1981,Weniger/1989}). It is not so well known among non-specialists
that some of these transformations sum many strongly divergent power
series much more effectively than Pad\'{e} approximants can do
it. Particularly suited for the summation of strongly divergent series
is a class of sequence transformations introduced by Levin
\cite{Levin/1973} in 1973. These transformations use as input data not
only the elements of a slowly convergent or divergent sequence $\{ s_n
\}_{n=0}^{\infty}$, whose elements may for instance be the partial sums
$s_n = \sum_{k=0}^{n} a_k$ of an infinite series, but also explicit
remainder estimates $\{ \omega_n \}_{n=0}^{\infty}$. Several
generalizations and extensions of Levin's transformation were derived
later, for instance in \cite[Sections 7 - 9]{Weniger/1989} or in
\cite{Weniger/1992}. Further details as well as the description of
several other Levin-type transformations can be found in a recent review
by Homeier \cite{Homeier/2000a}.  

The explicit incorporation of the information contained in the remainder
estimates $\{ \omega_n \}_{n=0}^{\infty}$ makes all Levin-type
transformation potentially very powerful (see for example the numerical
examples in \cite{Smith/Ford/1979,Smith/Ford/1982,Weniger/1989}). In the
case of divergent alternating series, the so-called delta transformation
\cite[Eq.\ (8.4-4)]{Weniger/1989} was found to be particularly useful
\cite{Borghi/Santarsiero/2003,Cizek/Vinette/Weniger/1991,
Cizek/Vinette/Weniger/1993,Cizek/Zamastil/Skala/2003,%
Jentschura/Becher/Weniger/Soff/2000,Jentschura/Weniger/Soff/2000,
Weniger/1989,Weniger/1990,Weniger/1992,Weniger/1994a,Weniger/1994b,
Weniger/1996a,Weniger/1996b,Weniger/1996c,Weniger/1996e,Weniger/1997,%
Weniger/2001,Weniger/Cizek/Vinette/1991,Weniger/Cizek/Vinette/1993}. 

However, sequence transformations in general or the Levin-type
transformations mentioned above in special are not only useful for the
summation of divergent perturbation expansions. In recent years, many
other successful applications of Levin-type transformations have been
reported in the literature (see for example
\cite{Aksenov/Savageau/Jentschura/Becher/Soff/Mohr/2003,%
Bar-Shalom/Klapisch/Oreg/1996,Bar-Shalom/Klapisch/Oreg/2001,Belkic/1989,%
Bhattacharya/Roy/Bhowmick/1997,Bhattacharya/Roy/Bhowmick/1989,%
Bouferguene/Fares/1994,Buesse/Kleindienst/Luechow/1998,%
Cizek/Vinette/Weniger/1991,Cizek/Vinette/Weniger/1993,DePrunele/1997,%
Dolovich/Brodland/1995,Edgal/Huber/1998,Grotendorst/1989,
Grotendorst/Steinborn/1986,Grotendorst/Weniger/Steinborn/1986,%
Homeier/1992,Homeier/1993,Homeier/2000a,Homeier/Weniger/1995,%
Ixaru/DeMeyer/VandenBerghe/2000,%
Jentschura/Gies/Valluri/Lamm/Weniger/2002,%
Jentschura/Mohr/Soff/1999,Jentschura/Mohr/Soff/2001a,%
Jentschura/Mohr/Soff/2001b,Jentschura/Mohr/Soff/Weniger/1999,%
Jetzke/Broad/1991,King/1999,King/Smethells/Helleloid/Pelzl/2002,%
Pelzl/King/1998,Pelzl/Smethells/King/2002,Oleksy/1996,Osada/1992a,%
Polly/Gruber/Windholz/Gleichmann/Hess/1998,Porras/King/1994,%
Roy/Bhattacharya/Bhowmick/1992,Roy/Bhattacharya/Bhowmick/1993,%
Roy/Bhattacharya/Bhowmick/1996,Roy/Bhattacharya/Bhowmick/1998,%
Sarkar/Haldar/Sen/Roy/2001,Sarkar/Sen/Haldar/Roy/1998,Schnack/2000,%
Scott/Aubert-Frecon/Andrae/2002,%
Soff/Bednyakov/Beier/Erler/Goidenko/Jentschura/et_al/2001,%
Steinborn/Weniger/1990,Veniard/Piraux/1991,Weniger/1989,Weniger/1990,%
Weniger/1992,Weniger/1994a,Weniger/1994b,Weniger/1996a,Weniger/1996b,%
Weniger/1996d,Weniger/2001,Weniger/Cizek/1990,%
Weniger/Grotendorst/Steinborn/1986a,Weniger/Steinborn/1989a}). This list
does not claim to be complete, but it suffices to show that Levin-type
transformations are extremely useful computational tools which deserve
to be more widely known.

In connection with the summation of the perturbation series for a
hydrogen atom in an external magnetic field, \v{C}\'{\i}\v{z}ek,
Zamastil, and Sk\'{a}la introduced a new sequence transformation
\cite[Eq.\ (10)]{Cizek/Zamastil/Skala/2003}, which in the notation of 
\cite{Weniger/1989} can be expressed as follows:
\begin{equation}
\mathcal{G}_{k}^{(n)} (q_m, s_n, \omega_n) \; = \; \frac
{\displaystyle
\sum_{j=0}^{k} \, (-1)^{j} \, {\binom{k}{j}} \,
\prod_{m=1}^{k-1} \, \frac {(n+j+q_m)}{(n+k+q_m)} \,
\frac {s_{n+j}} {\omega_{n+j}} }
{\displaystyle
\sum_{j=0}^{k} \, (-1)^{j} \, {\binom{k}{j}} \,
\prod_{m=1}^{k-1} \, \frac {(n+j+q_m)}{(n+k+q_m)} \,
\frac {1} {\omega_{n+j}} } 
\, , \qquad k, n \in \mathbb{N}_0 \, . 
\label{GenCiZaSkaTr}
\end{equation}
Here as well as later in the text it is always assumed that
$\prod_{k=l}^n a_k = 1$ holds if it is a so-called empty product with $l
> n$.

The sequence transformation (\ref{GenCiZaSkaTr}) contains the
unspecified parameters $q_m$ with $1 \le m \le k-1$. As discussed in
Section \ref{Sec:AnnihilOp} in more details, several other sequence
transformations can be obtained by specifying the parameters $q_m$. If
we for instance choose $q_m = \beta$ with $\beta > 0$, we obtain Levin's
transformation $\mathcal{L}_{k}^{(n)} (\beta, s_n, \omega_n)$
\cite{Levin/1973} in the notation of  \cite[7.1-7]{Weniger/1989}), if 
we choose $q_m = \beta + m - 1$, we obtain $\mathcal{S}_{k}^{(n)}
(\beta, s_n, \omega_n)$ \cite[Eq.\ (8.2-7)]{Weniger/1989}, which is the
parent transformation of the so-called delta transformation \cite[Eq.\
(8.4-4)]{Weniger/1989} mentioned above, and if we choose $q_m = \xi - m
+ 1$ with $\xi > 0$, we obtain $\mathcal{M}_{k}^{(n)} (\xi, s_n,
\omega_n)$ \cite[Eq.\ (9.2-6)]{Weniger/1989}. Then, there is a sequence
transformation $\mathcal{C}_{k}^{(n)} (\alpha, \beta, s_n,
\omega_n)$ \cite[Eq.\ (3.2)]{Weniger/1992} which -- depending on the
value of the parameter $\alpha$ -- interpolates between Levin's
transformation $\mathcal{L}_{k}^{(n)} (\beta, s_n, \omega_n)$ and
$\mathcal{S}_{k}^{(n)} (\beta, s_n, \omega_n)$. It is obtained by
choosing $q_m = \beta + [m-1]/\alpha$.

Thus, the transformation $\mathcal{G}_{k}^{(n)} ( q_m, s_n, \omega_n)$
introduced by \v{C}\'{\i}\v{z}ek, Zamastil, and Sk\'{a}la provides a
unifying concept for a large and practically important class of sequence
transformations, and all results derived for $\mathcal{G}_{k}^{(n)} (
q_m, s_n, \omega_n)$ can immediately be translated to the analogous
results for its various special cases. However, so far only the explicit
expression for this transformation as the ratio of two finite sums
according to (\ref{GenCiZaSkaTr}) is known
\cite[Eq.\ (10)]{Cizek/Zamastil/Skala/2003}, and many other
mathematical properties of interest are unknown. 

In Section \ref{Sec:AnnihilOp}, the explicit expression for
$\mathcal{G}_{k}^{(n)} ( q_m, s_n, \omega_n)$ is rederived by applying a
suitable annihilating difference operator to the model sequence
(\ref{CiZaSkaTrModSeq}) according to (\ref{GenSeqTr}). This annihilation
operator approach was originally introduced in
\cite[Section 3.2]{Weniger/1989} in connection with a simplified
derivation of the explicit expression for Levin's transformation
\cite{Levin/1973} and the construction of explicit expressions for 
other, closely related sequence transformations \cite[Sections 7 -
9]{Weniger/1989}. This annihilation operator approach does not only
produce the explicit expression (\ref{GenCiZaSkaTr}), but it also
provides a convenient starting point for the derivation of a recursive
scheme for the numerators and denominators in (\ref{GenCiZaSkaTr}),
which is done in Section \ref{Sec:RecForm}, and for a theoretical
convergence analysis, which will be done in \cite{Weniger/2003b*}.

In Section \ref{Sec:LevExpRemEst}, simple explicit remainder estimates
introduced by Levin \cite{Levin/1973} and Smith and Ford
\cite{Smith/Ford/1979}, which in the terminology of
\cite{Weniger/1989} yield the $u$, $t$, $v$, and $d$ variants of Levin's
sequence transformation, are used in combination with
$\mathcal{G}_{k}^{(n)} ( q_m, s_n, \omega_n)$. The effectivity of these
remainder estimates is motivated and studied via some model
sequences. Surprisingly, the $v$ type remainder estimate produces more
effective asymptotic estimates for the truncation errors of these model
sequences than the other simple remainder estimates. Moreover, it is
shown that all $t$ type variants considered in this article are actually
analogous $d$ type variants in disguise.

In Section \ref{Sec:RichTypeTrans}, variants of $\mathcal{G}_{k}^{(n)} (
q_m, s_n, \omega_n)$ are studied which closely resemble the Richardson
extrapolation process \cite{Richardson/1927}. These variants be used in
the case of logarithmic convergence ($\rho = 1$ in
(\ref{DefLinLogConv})), whose acceleration constitutes a formidable
computational problem.

In Section \ref{Sec:RatApprox}, the $u$, $t$, $d$, and $v$ variants of
$\mathcal{G}_{k}^{(n)} ( q_m, s_n, \omega_n)$ are applied to the partial
sums of a (formal) power series. This produces rational approximants
that resemble Pad\'{e} approximants, which are defined via the
accuracy-through-order relationship (\ref{O_est_Pade}). In the case of
the $u$, $t$, and $d$ variants, the resulting rational expressions are
actually Pad\'{e}-type approximants, which satisfy the modified
accuracy-through-order relationship (\ref{O_est_PadeType})
\cite{Brezinski/1980a}. In the case of the $v$ variant the resulting
expression is a slight generalization of a Pad\'{e}-type
approximant. With the help of the accuracy-through-order relationship
(\ref{O_est_PadeType}), which defines Pad\'{e}-type approximants, the
accuracy-through-order relationships satisfied by these rational
functions can be derived easily. These accuracy-through-order
relationships are needed if the rational approximants derived from
$\mathcal{G}_{k}^{(n)} ( q_m, s_n, \omega_n)$ are to be used for the
prediction of unknown power series coefficients.

This article is concluded in Section \ref{Sec:SummOutlook} by a short
summary and a critical assessment of the essential features of the new
Levin-type transformation introduced by recently \v{C}\'{\i}\v{z}ek,
Zamastil, and Sk\'{a}la \cite{Cizek/Zamastil/Skala/2003}.

Only the mathematical properties of $\mathcal{G}_{k}^{(n)} ( q_m, s_n,
\omega_n)$ and its various special cases are treated in this article,
albeit in a relatively detailed way. Anybody interested in other
sequence transformations should consult the monograph by Brezinski and
Redivo Zaglia \cite{Brezinski/RedivoZaglia/1991a}. It contains a wealth
of material and provides a very readable introduction to a rapidly
growing subfield of numerical mathematics. The older history of sequence
transformations up to about 1945 is treated in a monograph by Brezinski
\cite{Brezinski/1991a}, and the more recent history is discussed in two
articles, also by Brezinski \cite{Brezinski/1996,Brezinski/2000c}. 

Finally, one should not forget that the study of sequence
transformations remains incomplete without the simultaneous study of
Pad\'{e} approximants. Here, I recommend the book by Baker and
Graves-Morris \cite{Baker/Graves-Morris/1996}.

\typeout{==> Annihilation Operators}
\section{Explicit Expressions via Annihilation Operators}
\label{Sec:AnnihilOp}

Let us assume that $\{ s_n \}_{n=0}^{\infty}$ is a slowly convergent or
divergent sequence, whose elements may for instance be the partial sums
$s_n = \sum_{k=0}^{n} a_k$ of an infinite series. A \emph{sequence
transformation} is a rule which maps a sequence $\{ s_n
\}_{n=0}^{\infty}$ to a new sequence $\{ s'_n \}_{n=0}^{\infty}$ with
hopefully better numerical properties.

The basic step for the construction of a sequence transformation is the
assumption that the elements of a convergent or divergent sequence $\{
s_n \}_{n=0}^{\infty}$ can be partitioned into a (generalized) limit $s$
and a remainder $r_n$ according to
\begin{equation}
s_n \; = \; s + r_n \, , \qquad n \in \mathbb{N}_0 \, .
\label{s_n_r_n}
\end{equation}

A sequence transformation tries to accomplish an acceleration of
convergence or a summation by eliminating the remainders $r_n$ as
effectively as possible from the input data $s_n$ with the help of
numerical techniques. In realistic problems, a sequence transformation
can only eliminate approximations to the remainders. Consequently, the
transformed sequence $\{ s_n^{\prime} \}_{n=0}^{\infty}$ will also be of
the type of (\ref{s_n_r_n}), which means that $s^{\prime}_n$ can also be
partitioned into the (generalized) limit $s$ and a transformed remainder
$r_n^{\prime}$ according to
\begin{equation}
s_n^{\prime} \; = \; s + r_n^{\prime} \, ,
\qquad n \in \mathbb{N}_0 \, .
% \label{}
\end{equation}
The transformed remainders $\{ r_n^{\prime} \}_{n=0}^{\infty}$ are in
general different from zero for all finite values of $n$. However,
convergence is accelerated if the transformed remainders $\{
r_n^{\prime} \}_{n=0}^{\infty}$ vanish more rapidly than the original
remainders $\{ r_n \}_{n=0}^{\infty}$ according to
\begin{equation}
\lim_{n \to \infty} \, \frac {s^{\prime}_n - s} {s_n - s} \; = \;
\lim_{n \to \infty} \, \frac {r^{\prime}_n} {r_n} \; = \; 0 \, ,
\label{DefConvAcc}
\end{equation}
and a divergent sequence is summed if the transformed remainders $r'_n$
vanish as $n \to \infty$.

In practice, an in principle unlimited variety of different types of
remainders can occur. Therefore, it is essential to make some
assumptions -- either explicitly or implicitly -- which provide the
basis for the construction of a sequence transformation.

Let us assume that we have sufficient reason to believe that the
elements of a sequence $\{ s_n \}_{n=0}^{\infty}$ can for all $n \in
\mathbb{N}_0$ be expressed by an expansion of the following type:
\begin{equation}
s_n \; = \; s \, + \, \sum_{j=0}^{\infty} \, c_j \, \psi_j (n) \, .
\label{s_nSerExp}
\end{equation}
The $\psi_j (n)$ are assumed to be \emph{known} functions of $n$, but
otherwise essentially arbitrary, and the $c_j$ are unspecified
coefficients independent of $n$. Hence, the ansatz (\ref{s_nSerExp})
incorporates convergent as well as divergent sequences, depending upon
the behavior of the functions $\psi_j (n)$ as $n \to \infty$.

If we want to accelerate the convergence of $\{ s_n \}_{n=0}^{\infty}$
to its limit $s$ or to sum it in the case of divergence with the help of
a sequence transformation, we have to compute approximations to the
remainders $\sum_{j=0}^{\infty} c_j \psi_j (n)$ and to eliminate them
from the input data. However, the remainders of the sequence
(\ref{s_nSerExp}) contain an infinite number of unspecified coefficients
$c_j$. Consequently, a complete determination of the remainders and
their subsequent elimination cannot be accomplished by purely numerical
means.

Let us also assume that the functions $\{ \psi_{j} (n)
\}_{j=0}^{\infty}$ form an asymptotic sequence as $n \to \infty$, i.e, 
that they satisfy for all $j \in \mathbb{N}_0$
\begin{equation}
\psi_{j+1} (n) \; = \; \mathrm{o} \bigl(\psi_j (n) \bigr) \, ,
\qquad n \to \infty \, .
\label{psi_jAsySeq}
\end{equation}
The best, which a purely numerical process can accomplish, is the
elimination of a finite number of the leading terms of
(\ref{s_nSerExp}). Obviously, this corresponds to the transformation of
the sequence (\ref{s_nSerExp}) to a new sequence $\{ s'_n
\}_{n=0}^{\infty}$ whose elements satisfy
\begin{equation}
s'_n \; = \; s \, + \, \sum_{j=0}^{\infty} \, c'_j \,
\psi_{k+j} (n) \, ,
\qquad n \in \mathbb{N}_0 \, , \quad k \in \mathbb{N} \, .
\label{s'_nSerExp}
\end{equation}
The $c'_j$ are numerical coefficients that depend on $k$ as well as on
the coefficients $c_j$ in (\ref{s_nSerExp}).

Obviously, the original remainders $\sum_{j=0}^{\infty} c_j \psi_j (n)$
are not completely eliminated from the elements of the input sequence
(\ref{s_nSerExp}). Since, however, the functions $\{ \psi_{j} (n)
\}_{j=0}^{\infty}$ are by assumption an asymptotic sequence according to
(\ref{psi_jAsySeq}), the transformed remainders $\sum_{j=0}^{\infty}
c'_j \psi_{k+j} (n)$ in (\ref{s'_nSerExp}) should at least for
sufficiently large values of $k \in \mathbb{N}$ have significantly
better numerical properties than the original remainders
$\sum_{j=0}^{\infty} c_j \psi_j (n)$.

Assumptions about the $n$-dependence of the truncation errors $r_n$ can
be incorporated into the transformation process via model sequences. In
this approach, a sequence transformation is constructed which produces
the (generalized) limit $s$ of the model sequence
\begin{equation}
s_n \; = \; s \, + \, r_n \; = \; s
\, + \, \sum_{j=0}^{k-1} \, c_j \, \psi_j (n)  \, , 
\qquad k \in \mathbb{N} \, , \quad n \in \mathbb{N}_0 \, ,
\label{GenModSeq}
\end{equation}
if it is applied to $k+1$ consecutive elements $s_n$, $s_{n+1}$, \ldots,
$s_{n+k}$ of this model sequence. Since the $\psi_j (n)$ are assumed to
be known functions of $n$, an element of this model sequence contains
$k+1$ unknown, the (generalized) limit $s$ and the $k$ unspecified
coefficients $c_0$, $c_1$, \ldots, $c_{k-1}$. Accordingly, it follows
from Cramer's rule that a sequence transformation, which is exact for
the model sequence (\ref{GenModSeq}), can be expressed as the ratio of
two determinants (see for example \cite[Section
1.5]{Brezinski/RedivoZaglia/1991a}). However, determinantal
representations will not be considered here since they are
computationally unattractive. Fortunately, the sequence transformation,
which is exact for the general model sequence (\ref{GenModSeq}), can
also be computed recursively, as shown independently by Schneider
\cite{Schneider/1975}, H{\aa}vie \cite{Havie/1979} and Brezinski
\cite{Brezinski/1980b}. An alternative recursive scheme, which is more 
economical than the original recursive scheme, was later obtained by
Ford and Sidi \cite{Ford/Sidi/1987}.

A detailed discussion of the construction of sequence transformations
via model sequences as well as many examples can for instance be found
in the book by Brezinski and Redivo Zaglia
\cite{Brezinski/RedivoZaglia/1991a} or in \cite{Weniger/1989}.

Levin-type sequence transformations try to make the transformation
process more efficient by explicitly utilizing the information contained
in remainder estimates $\{ \omega_n \}_{n=0}^{\infty}$. Thus, a sequence
transformation is constructed which is exact for the elements of the
model sequence \cite[Eq.\ (3.2-9)]{Weniger/1989}
\begin{equation}
s_n \; = \; s \, + \, \omega_n z_n \, , \qquad n \in \mathbb{N}_0 \, .
\label{Mod_Seq_Om}
\end{equation}
The remainder estimates $\omega_n$ are assumed to be known, and the
correction terms $z_n$ should be chosen in such a way that the products
$\omega_n z_n$ provide sufficiently accurate and rapidly convergent
approximations to actual remainders. The principal advantage of this
approach is that only the correction terms $\{ z_n \}_{n=0}^{\infty}$
have to be determined. If good remainder estimates can be found, the
determination of $z_n$ and the subsequent elimination of $\omega_n z_n$
from $s_n$ often leads to substantially better results than the
construction and subsequent elimination of other approximations to
$r_n$.

The model sequence (\ref{Mod_Seq_Om}) has another indisputable
advantage: There exists a systematic approach for the construction of a
sequence transformation which is exact for this model sequence. Let us
assume that a \emph{linear} operator $\hat{T}$ can be found which
annihilates for all $n \in \mathbb{N}_0$ the correction term $z_n$
according to $\hat{T} (z_n) = 0$. Then, a sequence transformation, which
is exact for the model sequence (\ref{Mod_Seq_Om}), can be obtained by
applying $\hat{T}$ to $[s_n - s] / \omega_n = z_n$. Since $\hat{T}$
annihilates $z_n$ and is by assumption linear, the following sequence
transformation $\mathcal{T}$ is exact for the model sequence
(\ref{Mod_Seq_Om})
\cite[Eq.\ (3.2-11)]{Weniger/1989}:
\begin{equation}
\mathcal{T} (s_n, \omega_n) \; = \; \frac
{\hat{T} (s_n / \omega_n )} {\hat{T} (1 / \omega_n )} \; = \; s \, .
\label{GenSeqTr}
\end{equation}
The construction of sequence transformations via annihilation operators
was introduced in \cite[Section 3.2]{Weniger/1989} in connection with a
rederivation of Levin's transformation \cite{Levin/1973} and the
construction of some other, closely related sequence transformations
\cite[Sections 7 - 9]{Weniger/1989}. 

Later, this annihilation operator approach was discussed and extended in
books by Brezinski \cite{Brezinski/1997} and Brezinski and Redivo Zaglia
\cite{Brezinski/RedivoZaglia/1991a} and in articles by Brezinski
\cite{Brezinski/1996,Brezinski/1999c,Brezinski/2000b,Brezinski/2000c}, 
Brezinski and Matos \cite{Brezinski/Matos/1996}, Brezinski and Redivo
Zaglia \cite{Brezinski/RedivoZaglia/1994a,Brezinski/RedivoZaglia/1994b,%
Brezinski/RedivoZaglia/1996b}, Brezinski and Salam 
\cite{Brezinski/Salam/1995}, Homeier \cite{Homeier/2000a}, 
Matos \cite{Matos/2000}, and Weniger \cite{Weniger/1992}.

Simple and yet very powerful sequence transformations result
\cite[Sections 7 - 9]{Weniger/1989} if the annihilation operator 
$\hat{T}$ in (\ref{GenSeqTr}) is based upon the finite difference
operator $\Delta$ defined by $\Delta f (n) = f (n+1) - f (n)$. As is
well known, the $k$-th power of the finite difference operator
annihilates a polynomial $P_{k-1} (n)$ of degree $k - 1$ in $n$
according to $\Delta^k P_{k-1} (n) = 0$. Thus, the correction terms
$z_n$ in (\ref{Mod_Seq_Om}) should be chosen in such a way that
multiplication of $z_n$ by some suitable quantity $w_k (n)$ yields a
polynomial $P_{k-1} (n)$ of degree $k-1$ in $n$. If such a $w_k (n)$ can
be found, then
\begin{equation}
\Delta^k \, [w_k (n) \, z_n] \; = \; 
\Delta^k \, P_{k-1} (n) \; = \; 0 
% \label{}
\end{equation}
and the weighted difference operator $\hat{T} = \Delta^k w_k (n)$
annihilates $z_n$. Thus, the corresponding sequence transformation
(\ref{GenSeqTr}) is given by the ratio
\begin{equation}
\mathcal{T}_k^{(n)} ( w_k (n) \vert s_n, \omega_n )
\; = \; \frac
{\Delta^k [w_k (n) s_n / \omega_n]}
{\Delta^k [w_k (n) / \omega_n ]} \, .
\label{Seq_w_k}
\end{equation}

The sequence transformation (\ref{GenCiZaSkaTr}) introduced by
\v{C}\'{\i}\v{z}ek, Zamastil, and Sk\'{a}la can be constructed on the 
basis of the model sequence \cite[Eq.\ (9)]{Cizek/Zamastil/Skala/2003}
\begin{equation}
s_n \; = \; s \, + \,
\omega_n \, \sum_{j=0}^{k-1} \, \frac{c_j}{\prod_{m=1}^{j} (n+q_m)} \, ,
\qquad k \in \mathbb{N} \, , \quad n \in \mathbb{N}_0 \, .
\label{CiZaSkaTrModSeq}
\end{equation}
It will become clear later (compare (\ref{AsyCondRemEst}) and the
discussion related to it) that in (\ref{CiZaSkaTrModSeq}) as well as in
the model sequences (\ref{LevTrModSeq}), (\ref{WenTrModSeq}),
(\ref{MinTrModSeq}), and (\ref{CizTrModSeq}), which can be derived from
(\ref{CiZaSkaTrModSeq}), it makes sense to assume $c_0 \neq 0$.

Both in the model sequence (\ref{CiZaSkaTrModSeq}) as well as in the
sequence transformation, which is derived from it, we want to admit
$n=0$. Moreover, this model sequence should have a consistent behavior
for all $n \in \mathbb{N}_0$. Thus, we normally require $q_m > 0$ for $0
\le m \le k-1$, but otherwise these parameters are essentially
arbitrary. 

Multiplication of the sum in (\ref{CiZaSkaTrModSeq}) by
$\prod_{m=1}^{k-1} (n+q_m)$ yields $\sum_{j=0}^{k-1} c_j
\prod_{i=j+1}^{k-1} (n+q_m)$, which is a polynomial of degree $k-1$ in
$n$. Thus, $\hat{T} = \Delta^k \prod_{m=1}^{k-1} (n+q_m)$ is the
operator which annihilates the correction term in
(\ref{CiZaSkaTrModSeq}), and we obtain from (\ref{GenSeqTr}) the
following difference operator representation for the sequence
transformation introduced by \v{C}\'{\i}\v{z}ek, Zamastil, and Sk\'{a}la
\cite[Eq.\ (10)]{Cizek/Zamastil/Skala/2003}:
\begin{equation}
\mathcal{G}_{k}^{(n)} (q_m, s_n, \omega_n) \; = \; \frac
{\displaystyle 
\Delta^k \prod_{m=1}^{k-1} (n+q_m) \frac{s_n}{\omega_n}}
{\displaystyle 
\Delta^k \prod_{m=1}^{k-1} (n+q_m) \frac{1}{\omega_n}} \, ,
\qquad k, n \in \mathbb{N}_0 \, .
\label{CiZaSkaTrDiffOpRep}
\end{equation}
It will become clear later that this difference operator representation
is in some sense more fundamental and more important than the explicit
expression which can be derived easily from it. We only have to insert
the well-known relationship
\begin{equation}
\Delta^k f (n) \; = \; (-1)^k \, \sum_{j=0}^{k} \, (-1)^j {\binom{k}{j}}
\, f (n+j) 
\label{DeltaPk}
\end{equation}
into the numerator and denominator of (\ref{CiZaSkaTrDiffOpRep}) to
obtain
\begin{equation}
\mathcal{G}_{k}^{(n)} (q_m, s_n, \omega_n) \; = \; \frac
{\displaystyle
\sum_{j=0}^{k} \, (-1)^{j} \, {\binom{k}{j}} \,
\prod_{m=1}^{k-1} (n+j+q_m) \,
\frac {s_{n+j}} {\omega_{n+j}} }
{\displaystyle
\sum_{j=0}^{k} \, (-1)^{j} \, {\binom{k}{j}} \,
\prod_{m=1}^{k-1} (n+j+q_m) \,
\frac {1} {\omega_{n+j}} } 
\, , \qquad k, n \in \mathbb{N}_0 \, . 
\label{GenCiZaSkaTrUnnorm}
\end{equation}

In the case of large transformation orders $k$, terms that are large in
magnitude occur in the binomial sums in the numerator and denominator of
(\ref{GenCiZaSkaTrUnnorm}). The same problem occurs also in explicit
expressions for other Levin-type sequence transformation discussed
later. In the case of some FORTRAN compilers, this can easily lead to
OVERFLOW. To decrease the magnitude of the terms, it has become
customary to include an additional normalization factor. Thus, we divide
numerator and denominator of (\ref{GenCiZaSkaTrUnnorm}) by
$\prod_{m=1}^{k-1} (n+k+q_m)$ and obtain (\ref{GenCiZaSkaTr}). Such an
approach is always possible since the coefficients of a ratio like
(\ref{GenCiZaSkaTrUnnorm}) are only defined up to a common nonzero
factor.

If we choose in (\ref{CiZaSkaTrModSeq}) $q_m = \beta$ with $\beta > 0$,
we obtain the model sequence for Levin's transformation
\cite{Levin/1973} in the notation of \cite[Eq.\ (7.1-2)]{Weniger/1989}:
\begin{equation}
s_n \; = \; s \, + \,
\omega_n \, \sum_{j=0}^{k-1} \, \frac{c_j}{(\beta+n)^j} \, ,
\qquad k \in \mathbb{N} \, , \quad n \in \mathbb{N}_0 \, .
\label{LevTrModSeq}
\end{equation}
Thus, the corresponding annihilation operator is given by $\hat{T} =
\Delta^k (\beta+n)^{k-1}$ and Levin's transformation \cite{Levin/1973} 
can in the notation of \cite[Eq.\ (7.1-7)]{Weniger/1989} be expressed as
follows:
\begin{align}
\mathcal{L}_{k}^{(n)} (\beta, s_n, \omega_n) & \; = \;
\mathcal{G}_{k}^{(n)} (\beta, s_n, \omega_n) 
\; = \; \frac
{\Delta^k [(\beta+n)^{k-1} s_n / \omega_n]}
{\Delta^k [(\beta+n)^{k-1} / \omega_n]}
\label{LevTrDiffOpRep} \\
& \; = \; \frac
{\displaystyle
\sum_{j=0}^{k} \, (-1)^{j} \, {\binom{k}{j}} \,
\frac {(\beta + n +j )^{k-1}} {(\beta + n + k )^{k-1}} \,
\frac {s_{n+j}} {\omega_{n+j}} }
{\displaystyle
\sum_{j=0}^{k} \, (-1)^{j} \, {\binom{k}{j}} \,
\frac {(\beta + n +j )^{k-1}} {(\beta + n + k )^{k-1}} \,
\frac {1} {\omega_{n+j}} }
\, , \qquad k, n \in \mathbb{N}_0 \, .
\label{GenLevTr}
\end{align}

If we choose in (\ref{CiZaSkaTrModSeq}) $q_m = \beta + m - 1$ with
$\beta > 0$, we obtain the following model sequence \cite[Eq.\
(8.2-1)]{Weniger/1989} for the sequence transformation
$\mathcal{S}_k^{(n)} (\beta, s_n, \omega_n)$ which is a truncated
factorial series involving Pochhammer symbols $(\beta+n)_j = 
\Gamma(\beta+n+j)/\Gamma(\beta+n)$:
\begin{equation}
s_n \; = \; s \, + \,
\omega_n \, \sum_{j=0}^{k-1} \, \frac{c_j}{(\beta+n)_j} \, ,
\qquad k \in \mathbb{N} \, , \quad n \in \mathbb{N}_0 \, .
\label{WenTrModSeq}
\end{equation}
The fundamental properties of factorial series are for instance
discussed in books by Nielsen \cite{Nielsen/1965} and N\"orlund
\cite{Noerlund/1926,Noerlund/1954}. 

The annihilation operator for the model sequence (\ref{WenTrModSeq}) is
given by $\hat{T} = \Delta^k (\beta+n)_{k-1}$ and the sequence
transformation $\mathcal{S}_k^{(n)} (\beta, s_n, \omega_n)$, which is
exact for the model sequence (\ref{WenTrModSeq}), can be expressed as
follows \cite[Eq.\ (8.2-7)]{Weniger/1989}:
\begin{align}
\mathcal{S}_{k}^{(n)} (\beta, s_n, \omega_n) & \; = \;
\mathcal{G}_{k}^{(n)} (\beta + m - 1, s_n, \omega_n) 
\; = \; \frac
{\Delta^k [(\beta+n)_{k-1} s_n / \omega_n]}
{\Delta^k [(\beta+n)_{k-1} / \omega_n]}
\label{WenTrDiffOpRep} \\
& \; = \; \frac
{\displaystyle
\sum_{j=0}^{k} \, (-1)^{j} \, {\binom{k}{j}} \,
\frac {(\beta + n +j )_{k-1}} {(\beta + n + k )_{k-1}} \,
\frac {s_{n+j}} {\omega_{n+j}} }
{\displaystyle
\sum_{j=0}^{k} \, (-1)^{j} \, {\binom{k}{j}} \,
\frac {(\beta + n +j )_{k-1}} {(\beta + n + k )_{k-1}} \,
\frac {1} {\omega_{n+j}} }
\, , \qquad k, n \in \mathbb{N}_0 \, .
\label{GenWenTr}
\end{align}
The ratio (\ref{GenWenTr}) was originally derived by Sidi
\cite{Sidi/1981} for the construction of explicit expressions for
Pad\'{e} approximants of some special hypergeometric series. However,
Sidi's article \cite{Sidi/1981} provides no evidence that he intended to
use this ratio as a sequence transformation. Moreover, I am not aware of
any article of Sidi where the properties of the sequence transformation
$\mathcal{S}_{k}^{(n)} (\beta, s_n, \omega_n)$ were discussed or where
it was applied. Later, $\mathcal{S}_{k}^{(n)} (\beta, s_n, \omega_n)$
was apparently used in the master thesis of Shelef \cite{Shelef/1987}
for the numerical inversion of Laplace transforms, but it seems that the
results of this master thesis were not published elsewhere. The first
refereed and generally accessible article, where an application of
(\ref{GenWenTr}) as a sequence transformation was described, is
\cite{Weniger/Steinborn/1989a}, and the mathematical properties of 
(\ref{GenWenTr}) as a sequence transformation and in particular its
connection with factorial series were first discussed in
\cite[Section 8]{Weniger/1989}.

If we choose in (\ref{CiZaSkaTrModSeq}) $q_m = \xi - m + 1$ with $\xi >
0$, we obtain the following model sequence \cite[Eq.\
(9.2-1)]{Weniger/1989} for the sequence transformation
$\mathcal{M}_k^{(n)} (\xi, s_n, \omega_n)$:
\begin{align}
s_n & \; = \; s \, + \, \omega_n \, 
\sum_{j=0}^{k-1} \, \frac{c_j}{\prod_{m=1}^{j} (\xi+n-m+1)} \\
& \; = \; s \, + \,
\omega_n \, \sum_{j=0}^{k-1} \, \frac{(-1)^{j} c_j}{(-\xi-n)_j}
\; = \; s \, + \,
\omega_n \, \sum_{j=0}^{k-1} \, \frac{c'_j}{(-\xi-n)_j} \, ,
\qquad k \in \mathbb{N} \, , \quad n \in \mathbb{N}_0 \, .
\label{MinTrModSeq}
\end{align}
Thus, the corresponding annihilation operator is given by $\hat{T} =
\Delta^k (-\xi-n)_{k-1}$ and the sequence transformation
$\mathcal{M}_k^{(n)} (\xi, s_n, \omega_n)$ can be expressed as
follows \cite[Eq.\ (9.2-6)]{Weniger/1989}:
\begin{align}
\mathcal{M}_{k}^{(n)} (\xi, s_n, \omega_n) & \; = \;
\mathcal{G}_{k}^{(n)} (\xi - m + 1, s_n, \omega_n) 
\; = \; \frac
{\Delta^k [(-\xi-n)_{k-1} s_n / \omega_n]}
{\Delta^k [(-\xi-n)_{k-1} / \omega_n]} 
\label{MinTrDiffOpRep} \\
& \; = \; \frac
{\displaystyle
\sum_{j=0}^{k} \, (-1)^{j} \, {\binom{k}{j}} \,
\frac {(-\xi-n-j )_{k-1}} {(-\xi-n-k )_{k-1}} \,
\frac {s_{n+j}} {\omega_{n+j}} }
{\displaystyle
\sum_{j=0}^{k} \, (-1)^{j} \, {\binom{k}{j}} \,
\frac {(-\xi-n-j )_{k-1}} {(-\xi-n-k )_{k-1}} \,
\frac {1} {\omega_{n+j}} }
\, , \qquad k, n \in \mathbb{N}_0 \, .
\label{GenMinTr}
\end{align}

If we choose in (\ref{CiZaSkaTrModSeq}) $q_m = \beta + [m-1]/\alpha$
with $\alpha, \beta > 0$, we obtain the following model sequence
\cite[Eq.\ (3.1)]{Weniger/1992} for the sequence transformation
$\mathcal{C}_k^{(n)} (\alpha, \beta, s_n, \omega_n)$:
\begin{align}
s_n & \; = \; s \, + \, \omega_n \, \sum_{j=0}^{k-1} \, 
\frac{c_j}{\prod_{m=1}^{j} (\beta+n+[m-1]/\alpha)} \\
& \; = \; s \, + \, \omega_n \, 
\sum_{j=0}^{k-1} \, \frac{\alpha^{j} c_j}{(\alpha[\beta+n])_j} 
\; = \; s \, + \, \omega_n \, 
\sum_{j=0}^{k-1} \, \frac{c'_j}{(\alpha[\beta+n])_j} \, ,
\qquad k \in \mathbb{N} \, , \quad n \in \mathbb{N}_0 \, .
\label{CizTrModSeq}
\end{align}
Thus, the corresponding annihilation operator is given by $\hat{T} =
\Delta^k (\alpha[\beta+n])_{k-1}$ and the sequence transformation
$\mathcal{C}_k^{(n)} (\alpha, \beta, s_n, \omega_n)$ can be expressed as
follows \cite[Eq.\ (3.2)]{Weniger/1992}:
\begin{align}
\mathcal{C}_{k}^{(n)} (\alpha, \beta, s_n, \omega_n) & \; = \;
\mathcal{G}_{k}^{(n)} (\beta + [m-1]/\alpha, s_n, \omega_n) 
\; = \; \frac
{\Delta^k [(\alpha[\beta+n])_{k-1} s_n / \omega_n]}
{\Delta^k [(\alpha[\beta+n])_{k-1} / \omega_n]}
\label{CizTrDiffOpRep} \\
& \; = \; \frac
{\displaystyle
\sum_{j=0}^{k} \, (-1)^{j} \, {\binom{k}{j}} \,
\frac
{(\alpha[\beta + n + j])_{k-1}} {(\alpha[\beta + n + k])_{k-1}} \,
\frac {s_{n+j}} {\omega_{n+j}} }
{\displaystyle
\sum_{j=0}^{k} \, (-1)^{j} \, {\binom{k}{j}} \,
\frac
{(\alpha[\beta + n + j])_{k-1}} {(\alpha[\beta + n + k])_{k-1}} \,
\frac {1} {\omega_{n+j}}}
\, , \qquad k, n \in \mathbb{N}_0 \, .
\label{GenCizTr}
\end{align}

Depending upon the value of $\alpha > 0$, the sequence transformation
$\mathcal{C}_{k}^{(n)} (\alpha, \beta, s_n, \omega_n)$ interpolates
between $\mathcal{S}_{k}^{(n)} (\beta, s_n, \omega_n)$ and Levin's
sequence transformation $\mathcal{L}_{k}^{(n)} (\beta, s_n, \omega_n)$.
If we choose $\alpha = 1$ in (\ref{GenCizTr}) and compare the resulting
expression with (\ref{GenWenTr}), we find
\begin{equation}
\mathcal{C}_{k}^{(n)} (1, \beta, s_n, \omega_n) \; = \;
\mathcal{S}_{k}^{(n)} (\beta, s_n, \omega_n) \, ,
% \label{}
\end{equation}
and if we use 
\begin{equation}
\lim_{\alpha \to \infty} \, 
\frac{(\alpha[\beta + n + j])_{k-1}}{\alpha^{k-1}} 
\; = \; \lim_{\alpha \to \infty} \, 
\prod_{m=1}^{k-1} \, (\beta + n + j + [m-1]/\alpha)
\; = \; (\beta + n + j)^{k-1}
% \label{}
\end{equation}
in (\ref{GenCizTr}) and compare the resulting expression with
(\ref{GenLevTr}), we find 
\begin{equation}
\lim_{\alpha \to \infty} \, 
\mathcal{C}_{k}^{(n)} (\alpha, \beta, s_n, \omega_n) \; = \;
\mathcal{L}_{k}^{(n)} (\beta, s_n, \omega_n) \, .
% \label{}
\end{equation}

Thus, the construction of explicit expressions for the Levin-type
sequence transformations $\mathcal{G}_{k}^{(n)} ( q_m, s_n, \omega_n)$,
$\mathcal{L}_{k}^{(n)} (\beta, s_n, \omega_n)$, $\mathcal{S}_{k}^{(n)}
(\beta, s_n, \omega_n)$, $\mathcal{M}_{k}^{(n)} (\xi, s_n, \omega_n)$,
and $\mathcal{C}_{k}^{(n)} (\alpha, \beta, s_n, \omega_n)$ with the help
of annihilation operators is almost trivial. However, the annihilation
operator approach does not only work in the case of the comparatively
simple annihilating difference operators in (\ref{CiZaSkaTrDiffOpRep}),
(\ref{LevTrDiffOpRep}), (\ref{WenTrDiffOpRep}), (\ref{MinTrDiffOpRep}),
and (\ref{CizTrDiffOpRep}). In \cite[Section 7.4]{Weniger/1989} it was
shown that Richardson extrapolation \cite{Richardson/1927} and Sidi's
generalized Richardson extrapolation process \cite{Sidi/1982d} can be
derived by using divided differences as annihilation operators. Then, it
was shown by Brezinski and Redivo Zaglia
\cite{Brezinski/RedivoZaglia/1994a,Brezinski/RedivoZaglia/1994b}, 
Brezinski and Matos \cite{Brezinski/Matos/1996}, and Matos
\cite{Matos/2000} that the majority of the currently known
transformations for scalar sequence can be derived via the annihilation
operator approach. The generalization of this approach to vector and
matrix sequences was discussed by Brezinski and Redivo Zaglia
\cite{Brezinski/RedivoZaglia/1996b} and Brezinski and Salam
\cite{Brezinski/Salam/1995}. 

\typeout{==> Recurrence Formulas}
\section{Recurrence Formulas}
\label{Sec:RecForm}

In the theory of sequence transformations it is relatively uncommon that
closed form expressions of the type of (\ref{GenCiZaSkaTr}),
(\ref{GenLevTr}), (\ref{GenWenTr}), (\ref{GenMinTr}), and
(\ref{GenCizTr}) are known. The majority of the currently known sequence
transformations are defined and computed via recursive schemes. From a
computational point of view, the lack of an explicit expression is
normally no disadvantage. The use of recurrence formulas is in most
cases (much) much more efficient, in particular if a whole sequence of
transforms must be computed simultaneously.

It is an additional advantage of the annihilation operator approach
described in Section \ref{Sec:AnnihilOp} that it permits a convenient
construction of a recursive scheme for the numerators and denominators
of $\mathcal{G}_{k}^{(n)} ( q_m, s_n, \omega_n)$. For that purpose, let
us define
\begin{align}
\Gamma_{k}^{(n)} & \; = \; \Gamma_{k}^{(n)} (q_m, u_n) \; = \;
\Delta^k X_{k}^{(n)} \, , \qquad k, n \in \mathbb{N}_0 \, , 
\label{DefGamma_kn} \\
X_{k}^{(n)} & \; = \; X_{k}^{(n)} (q_m, u_n) \; = \;
\prod_{m=1}^{k-1} (n+q_m) \, u_n \, , \qquad k, n \in \mathbb{N}_0 \, .
\label{DefXkn}
\end{align}
Comparison with (\ref{CiZaSkaTrDiffOpRep}) shows that $\Gamma_{k}^{(n)}$
corresponds apart from a missing phase factor $(-1)^k$ to the numerator
in (\ref{GenCiZaSkaTrUnnorm}) if we choose $u_n = s_n/\omega_n$, and to
the denominator in (\ref{GenCiZaSkaTrUnnorm}) if we choose $u_n =
1/\omega_n$.

The quantities $X_{k}^{(n)}$ satisfy for $k \ge 2$ the two-term
recursion
\begin{equation}
X_{k}^{(n)} \; = \; (n+q_{k-1}) \, X_{k-1}^{(n)} \, .
\label{RecXkn}
\end{equation}
Next, we use the commutator relationship (see
\cite{Fessler/Ford/Smith/1983a} or \cite[Eq.\ (7.2-2)]{Weniger/1989})
\begin{equation}
\Delta^k \, (n+q_{k-1}) \, - \, (n+q_{k-1}) \, \Delta^k \; = \;
k \, E \, \Delta^{k-1} \, ,
\label{Comm1}
\end{equation}
which can be proved by complete induction. This commutator can be
rewritten as follows:
\begin{align}
\Delta^k \, (n+q_{k-1}) & \; = \; (n+k+q_{k-1}) \, \Delta^k \, + \, 
k \, \Delta^{k-1} 
\label{comm2} 
\\
& \; = \; \bigl[ (n+k+q_{k-1}) \, (E-1) \, + \, k \bigr] 
\, \Delta^{k-1} 
\label{comm3} 
\\
& \; = \; \bigl[ (n+k+q_{k-1}) \, E \, - \, (n+q_{k-1}) \bigr] 
\, \Delta^{k-1} \, . 
\label{comm4}
\end{align}
Here, $E$ is the shift operator defined by $E f (n) = f (n+1)$.

Combination of (\ref{DefGamma_kn}), (\ref{DefXkn}), (\ref{RecXkn}) and
(\ref{comm4}) yields:
\begin{align}
\Gamma_{k}^{(n)} & \; = \; \Delta^k \, (n+q_{k-1}) \, X_{k-1}^{(n)} \\
& \; = \; \bigl[ (n+k+q_{k-1}) \, E \, - \, (n+q_{k-1}) \bigr] 
\, \Delta^{k-1}  \, X_{k-1}^{(n)} \\
& \; = \; (n+k+q_{k-1}) \, \Gamma_{k-1}^{(n+1)} \, - \, 
(n+q_{k-1}) \, \Gamma_{k-1}^{(n)} \, , 
\qquad k \ge 2 \, , \quad n \in \mathbb{N}_0 \, .
\label{RecGamma_kn}
\end{align}
Finally, we rescale $\Gamma_{k}^{(n)}$ according to
\begin{equation}
G_{k}^{(n)} \; = \;
G_{k}^{(n)} (q_m, u_n) \; = \; \frac
{\Gamma_{k}^{(n)} (q_m, u_n)}{\prod_{m=1}^{k-1} (n+k+q_m)} \, ,
\qquad k, n \in \mathbb{N}_0 \, .
\label{DefGkn}
\end{equation}
If we combine (\ref{RecGamma_kn}) and (\ref{DefGkn}) and take into
account that (\ref{DefGamma_kn}) and (\ref{DefXkn}) imply
$\Gamma_{0}^{(n)} = u_n$ and $\Gamma_{1}^{(n)} = u_{n+1} - u_n$,
respectively, we obtain the following recursive scheme for the
numerators and denominators of the sequence transformation
(\ref{GenCiZaSkaTr}) introduced by \v{C}\'{\i}\v{z}ek, Zamastil, and
Sk\'{a}la \cite[Eq.\ (10)]{Cizek/Zamastil/Skala/2003}:
\begin{subequations}
\label{RecSchemeCiZaSkaTr}
\begin{align}
G_{0}^{(n)} & \; = \; u_n \, , 
\quad n \in \mathbb{N}_0 \, ,
\label{RecSchemeCiZaSkaTr_a} 
\\
G_{1}^{(n)} & \; = \; u_{n+1} \, - \, u_n \, , 
\quad n \in \mathbb{N}_0 \, ,
\label{RecSchemeCiZaSkaTr_b} 
\\
G_{k+1}^{(n)} & \; = \; G_{k}^{(n+1)} \, - \, 
\frac{n+q_{k}}{n+k+q_{k}+1} \, 
\prod_{m=1}^{k-1} \, \frac{n+k+q_m}{n+k+q_m+1} \,
G_{k}^{(n)} \, , 
\quad k  \in \mathbb{N} \, , \quad n \in \mathbb{N}_0 \, .
\label{RecSchemeCiZaSkaTr_c}
\end{align}
\end{subequations}
If we choose $u_n = s_n/\omega_n$, this recursive scheme produces the
numerator of (\ref{GenCiZaSkaTr}), and if we choose $u_n = 1/\omega_n$,
we obtain the denominator of (\ref{GenCiZaSkaTr}).

As shown in \cite[Sections 7.3, 8.3, and 9.3]{Weniger/1989}, recursive
schemes for the numerator and denominator sums of the sequence
transformations $\mathcal{L}_{k}^{(n)} (\beta, s_n,
\omega_n)$. $\mathcal{S}_{k}^{(n)} (\beta, s_n, \omega_n)$, and
$\mathcal{M}_{k}^{(n)} (\xi, s_n, \omega_n)$ can be derived in the same
way. The recursive scheme \cite[Eq.\ (7.2-8))]{Weniger/1989}
\begin{subequations}
\label{RecSchemeLevTr}
\begin{align}
L_{0}^{(n)} & \; = \; u_n \, , 
\quad n \in \mathbb{N}_0 \, , \\
L_{k+1}^{(n)} & \; = \; L_k^{(n+1)} \, - \, \frac
{(\beta + n) (\beta + n + k)^{k-1}}{(\beta + n + k + 1)^k} \,
L_k^{(n)} \, , \quad k, n \in \mathbb{N}_0 \, ,
\end{align}
\end{subequations}
produces the numerator and denominator sums in (\ref{GenLevTr}) if we
choose $u_n = s_n/\omega_n$ and $u_n = 1/\omega_n$, respectively. We
obtain (\ref{RecSchemeLevTr}) from (\ref{RecSchemeCiZaSkaTr}) by
choosing $q_m = \beta$.

Similarly, the recursive schemes \cite[Eq.\ (8.3-7))]{Weniger/1989}
\begin{subequations}
\label{RecSchemeWenTr}
\begin{align}
S_{0}^{(n)} & \; = \; u_n \, , 
\quad n \in \mathbb{N}_0 \, , \\
S_{k+1}^{(n)} & \; = \; S_k^{(n+1)} \, - \, \frac
{(\beta + n + k - 1) (\beta + n + k)}
{(\beta + n + 2k - 1) (\beta + n + 2k)} \,
S_k^{(n)} \, , \quad k, n \in \mathbb{N}_0 \, ,
\end{align}
\end{subequations}
and \cite[Eq.\ (9.3-6))]{Weniger/1989}
\begin{subequations}
\label{RecSchemeMinTr}
\begin{align}
M_{0}^{(n)} & \; = \; u_n \, , 
\quad n \in \mathbb{N}_0 \, , \\
M_{k+1}^{(n)} & \; = \; M_k^{(n+1)} \, - \, \frac
{\xi + n - k + 1}{\xi + n + k + 1} \,
M_k^{(n)} \, , \quad k, n \in \mathbb{N}_0 \, ,
\end{align}
\end{subequations}
produce the numerator and denominator sums in (\ref{GenWenTr}) and
(\ref{GenMinTr}), respectively. The recursive schemes
(\ref{RecSchemeWenTr}) and (\ref{RecSchemeMinTr}) can be obtained from
(\ref{RecSchemeCiZaSkaTr}) by setting $q_m = \beta + m - 1$ and $q_m =
\xi - m + 1$, respectively.

If we set $q_m = \beta + [m-1]/\alpha$ in (\ref{RecSchemeCiZaSkaTr}), we
obtain the recursive scheme for the interpolating transformation
$\mathcal{C}_{k}^{(n)} (\alpha, \beta, s_n, \omega_n)$:
\begin{subequations}
\label{RecSchemeCizTr}
\begin{align}
C_{0}^{(n)} & \; = \; u_n \, , 
\quad n \in \mathbb{N}_0 \, , \\
C_{1}^{(n)} & \; = \; u_{n+1} \, - \, u_n \, , 
\quad n \in \mathbb{N}_0 \, , \\
C_{k+1}^{(n)} & \; = \; C_k^{(n+1)} \, - \, \frac
{(\alpha[\beta + n] + k - 1) (\alpha[\beta + n + k])_{k-1}}
{(\alpha[\beta + n + k + 1])_k} \, C_k^{(n)} \, , 
\quad k \in \mathbb{N} \, , n \in \mathbb{N}_0 \, .
\end{align}
\end{subequations}
This scheme produces the numerator and denominator sums of
(\ref{GenCizTr}) if we choose $u_n = s_n/\omega_n$ and $u_n =
1/\omega_n$, respectively. The recurrence formula published in
\cite[Eq.\ (3.3)]{Weniger/1992} contains errors.

\typeout{==> Levin's Explicit Remainder Estimates}
\section{Levin's Explicit Remainder Estimates}
\label{Sec:LevExpRemEst}

It follows from (\ref{CiZaSkaTrDiffOpRep}), (\ref{DeltaPk}),
(\ref{LevTrDiffOpRep}), (\ref{WenTrDiffOpRep}), (\ref{MinTrDiffOpRep}),
and (\ref{CizTrDiffOpRep}) that all sequence transformations considered
in this article can be expressed as follows:
\begin{equation}
T_{k}^{(n)} (s_n, \omega_n) \; = \; \frac
{\displaystyle \Delta^k \, \{ P_{k-1} (n) \, s_n/\omega_n \} }
{\displaystyle \Delta^k \, \{ P_{k-1} (n)/\omega_n \} } \, ,
\qquad k \in \mathbb{N} \, , \quad n \in \mathbb{N}_0 \, .
\label{GenLevTypeTrDiffOpRep}
\end{equation}
Here, $P_{k-1} (n)$ is a polynomial of degree $k-1$ in $n$. Obviously,
the remainder estimates $\{ \omega_n \}_{n=0}^{\infty}$ have to satisfy
the minimal requirement that $\Delta^k \{ P_{k-1} (n)/\omega_n \} \ne 0$
for all \emph{finite} $k, n \in \mathbb{N}_0$. In the following text,
this will always be assumed.

The weighted difference operator $\Delta^k P_{k-1} (n)$ in
(\ref{GenLevTypeTrDiffOpRep}) is linear. Accordingly, such a sequence
transformation satisfies
\begin{equation}
T_{k}^{(n)} (s_n, \omega_n) \; = \; s \, + \, \frac
{\displaystyle 
\Delta^k \, \bigl\{ P_{k-1} (n) \, [s_n - s]/\omega_n\bigr\} }
{\displaystyle 
\Delta^k \, \bigl\{ P_{k-1} (n)/\omega_n\bigr\} } \, ,
\qquad k \in \mathbb{N} \, , \quad n \in \mathbb{N}_0 \, .
\label{GenLevTypeTrDiffOpRepLimPlusRn}
\end{equation}
This property of Levin-type transformations has some far-reaching
consequences. Let us assume that we can find for a given sequence $\{
s_n \}_{n=0}^{\infty}$ a sequence $\{ \omega_n \}_{n=0}^{\infty}$ of 
\emph{perfect} remainder estimates such that
\begin{equation}
s_n \; = \; s \, + \, r_n \; = \; s \, + \, c \, \omega_n \, , 
\qquad c \ne 0 \, , \quad n \in \mathbb{N}_0 \, .
\label{SeqPerfecRemEst}
\end{equation}
The polynomial $P_{k-1} (n)$ in (\ref{GenLevTypeTrDiffOpRep}) is of
degree $k-1$ in $n$, which implies that it is annihilated by $\Delta^k$
according to $\Delta^k P_{k-1} (n) = 0$. Accordingly, the transformation
problem is now trivial since (\ref{GenLevTypeTrDiffOpRepLimPlusRn})
produces the (generalized) limit $s$ of the sequence
(\ref{SeqPerfecRemEst}) \cite[Theorem 12-8]{Weniger/1989}):
\begin{equation}
T_{k}^{(n)} (s_n, \omega_n) \; = \; s \, + \, \frac
{c \Delta^k P_{k-1} (n)}
{\Delta^k \bigl\{ P_{k-1} (n)/\omega_n \bigr\} } \; = \; s \, , 
\qquad k \in \mathbb{N} \, , \quad n \in \mathbb{N}_0 \, .
% \label{}
\end{equation}

Unfortunately, perfect remainder estimates satisfying
(\ref{SeqPerfecRemEst}) can only be found for practically more or less
irrelevant model problems. In the case of realistic problems, we have to
be more modest and can only hope to find remainder estimates that
reproduce the leading order asymptotics of the actual remainders
\cite[Eq.\ (7.3-1)]{Weniger/1989}:
\begin{equation}
r_n \; = \; s_n \, - \, s \; = \; \omega_n \, 
\left[ c + \mathrm{O} \bigl( 1/n \bigr) \right] \, ,
\qquad c \ne 0 \, , \quad n \to \infty \, .
\label{AsyCondRemEst}
\end{equation} 
This asymptotic condition does not fix remainder estimates uniquely. All
Levin-type transformations considered in this article are invariant
under the transformation $\omega_n \to c \omega_n$ with $c \ne
0$. Moreover, given a sequence $\{ r_n \}_{n=0}^{\infty}$ of remainders
it is usually possible to find a variety of genuinely different
sequences $\{ \omega_n \}_{n=0}^{\infty}$, $\{ \omega'_n
\}_{n=0}^{\infty}$, $\{ \omega''_n \}_{n=0}^{\infty}$, \ldots\ of 
remainder estimates which all satisfy the asymptotic condition
(\ref{AsyCondRemEst}).

In some exceptional cases, explicit analytical expressions for remainder
estimates can be found. Let us for instance assume that the elements of
the sequence to be transformed are the partial sums $s_n =
\sum_{k=0}^{n} a_k$ of an infinite series and that the terms $a_k$ have
a sufficiently simple analytical structure. Then it may be possible to
derive an explicit expression for the truncation error, from which
explicit remainder estimates satisfying (\ref{AsyCondRemEst}) can be
derived.

In principle, such an analytical approach would be highly desirable, in
particular since it should then be possible to construct for a given
sequence $\{ s_n \}_{n=0}^{\infty}$ more effective remainder estimates
$\bigl\{ \tilde{\omega}_n^{(l)} \bigr\}_{n=0}^{\infty}$ that do not only
reproduce the leading order asymptotics of the remainders according to
(\ref{AsyCondRemEst}), but several of the leading orders according to
\begin{equation}
r_n \; = \; s_n \, - \, s \; = \; \tilde{\omega}_n^{(l)} \, 
\left[ c + \mathrm{O} \bigl( n^{-l} \bigr) \right] \, ,
\quad c \ne 0 \, , \quad n \to \infty \, ,
\label{ModAsyCondRemEst}
\end{equation} 
where $l > 1$ is a fixed positive integer. Improved remainder estimates
of that kind should lead to more efficient Levin-type
transformations. Unfortunately, only relatively little work has been
done on the asymptotics of truncation errors $\sum_{k=n+1}^{\infty} a_k$
of infinite series as $n \to \infty$ beyond the leading order. Moreover,
in many applications of Levin-type transformations in particular in
physics, only the numerical values of a finite string of sequence
elements or series coefficients are available, but no explicit
analytical expressions. In such a case, remainder estimates have to be
constructed from the numerical values of the input data via simple
rules.

Levin-type sequence transformations are not limited to strongly
divergent perturbation expansions. They are able to accelerate the
convergence of many series and sequences if suitable remainder estimates
are used. In this context, it is helpful to introduce first some
terminology which is common in the literature on convergence acceleration
methods. Many practically relevant sequences $\{ s_n
\}_{n=0}^{\infty}$, which converge to some limit $s$, satisfy
\begin{equation}
\lim_{n \to \infty} \; \frac {s_{n+1} - s} {s_n - s} \; = \; \lim_{n
\to \infty} \; \frac {r_{n+1}}{r_n} \; =\; \rho \, .
\label{DefLinLogConv}
\end{equation}
If $0 < \vert \rho \vert < 1$ holds, we say that the sequence $\{ s_n
\}_{n=0}^{\infty}$ converges \emph{linearly}, if $\rho = 1$ holds, we say
that this sequence converges \emph{logarithmically}, and if $\rho = 0$
holds, we say that it converges \emph{hyperlinearly}. Of course, $\vert
\rho \vert > 1$ implies that the sequence $\{ s_n \}_{n=0}^{\infty}$
diverges.

On the basis of purely heuristic arguments Levin \cite{Levin/1973} and
later Smith and Ford \cite{Smith/Ford/1979} suggested some simple
remainder estimates which according to experience nevertheless work
remarkably well in a large variety of cases. These simple remainder
estimates can be motivated by considering simple model problems. For
that purpose, let us assume that the elements of a sequence $\{ s_n
\}_{n=0}^{\infty}$ of partial sums $\sum_{\nu=0}^{n} a_{\nu}$ behave as
follows:
\begin{equation}
s_n \; \sim \; s \, + \, z^{n+1} \, n^{\theta} \, 
\left[ \alpha_0 + \frac{\alpha_{1}}{n} + \frac{\alpha_{2}}{n^2} 
\, + \, \ldots \right] \, , \qquad n \to \infty \, .
\label{AsyLinLogModSeq}
\end{equation}
This is a fairly general model sequence, which is able to describe the
asymptotics of many practically relevant sequences as $n \to \infty$ and
which Levin \cite{Levin/1973} had in mind when he introduced his simple
remainder estimates. For $\vert z \vert < 1$, the sequence
(\ref{AsyLinLogModSeq}) converges linearly to its limit $s$, for $z = 1$
and $\mathrm{Re} (\theta) < 0$, it converges logarithmically, and for
$\vert z \vert > 1$ it diverges.

From (\ref{AsyLinLogModSeq}) we obtain via $a_n = \Delta s_{n-1}$ the
leading orders of the asymptotic expansion of the terms of the infinite
series $\sum_{\nu=0}^{\infty} a_{\nu}$. For $z = 1$ and $\mathrm{Re}
(\theta) < 0$ (logarithmic convergence), we find with the help of the
computer algebra system Maple
\begin{eqnarray}
a_n & = &  n^{\theta} \, \biggl\{ \frac{\theta \alpha_0}{n} +  
\frac{(\theta-1) [2 \alpha_1 - \theta \alpha_0]}{2 n^2} \nonumber \\
& & \quad + \, 
{\displaystyle \frac{(\theta-2) \left[ 6 \alpha_2 + (\theta-1)
\{ \theta \alpha_0 - 3 \alpha_1 \} \right]}{6 n^3}} 
\, + \, \mathrm{O} \bigl( n^{-3} \bigr) \biggr\} \, , 
\qquad n \to \infty \, ,
\label{TermLogConv}
\end{eqnarray}
and for $\vert z \vert < 1$ (linear convergence), we find.
\begin{eqnarray}
a_n & = & z^n \, n^{\theta} \, \biggl\{ (z-1) \alpha_0 + 
\frac{\theta \alpha_0 + (z-1) \alpha_1}{n} \nonumber \\
&& \qquad + \, 
{\displaystyle \frac{(\theta-1) \left[2 \alpha_1 - \theta
\alpha_0 \right] + 2 (z-1) \alpha_2}{2 n^2}}
\, + \, \mathrm{O} \bigl( n^{-3} \bigr) \biggr\} \, , 
\qquad n \to \infty \, .
\label{TermLinConv}
\end{eqnarray}

If we compare (\ref{AsyLinLogModSeq}) and (\ref{TermLogConv}), we see
that in the case of logarithmic convergence the term $a_n = \Delta
s_{n-1} = \mathrm{O} (n^{\theta-1})$ cannot reproduce the leading order
of the remainder $r_n = s_n - s = \mathrm{O} (n^{\theta})$. However, the
product $n a_n = \mathrm{O} (n^{\theta})$ reproduces the leading order
of the remainder of the model sequence (\ref{AsyLinLogModSeq}). Thus,
it is an obvious idea to use the remainder estimate \cite{Levin/1973}
\begin{equation}
\omega_n \; = \; (\beta + n) \Delta s_{n-1} \; = \; (\beta + n) a_n
\label{uRemEst}
\end{equation}
in Levin's general transformation (\ref{GenLevTr}), yielding Levin's $u$
transformation in the notation of \cite[Eq.\ (7.3-5)]{Weniger/1989}: 
\begin{equation}
u_{k}^{(n)} (\beta, s_n) \; = \;
\mathcal{L}_{k}^{(n)} (\beta, s_n, (\beta + n) \Delta s_{n-1}) 
\, , \qquad k, n \in \mathbb{N}_0 \, .
\label{uLevTr}
\end{equation}

Since Levin's remainder estimate (\ref{uRemEst}) reproduces the leading
order of the remainder of the model sequence (\ref{AsyLinLogModSeq}) for
$z = 1$ and $\mathrm{Re} (\theta) < 0$ (logarithmic convergence), it is
not surprising that the $u$ transformation is an effective accelerator
for many monotone, logarithmically convergent sequences and series.

In the case of linear convergence ($\vert z \vert < 1$ in
(\ref{AsyLinLogModSeq})), the asymptotic expansion (\ref{TermLinConv})
indicates that the term $a_n$ itself and not the product $n a_n$ would
be a natural estimate for the truncation error in
(\ref{AsyLinLogModSeq}). However, the Levin-type transformations
considered in this article nevertheless accelerate convergence if
instead of the ``right'' sequence $\{ \omega_n \}_{n=0}^{\infty}$ of
remainder estimates ``wrong'' remainder estimates $\omega'_n =
(n+\beta)^l \omega_n$ with $l \in \mathbb{N}_0$ are used (see Theorems
12-14 - 12-16 and the discussion on pp.\ 310 - 311 of
\cite{Weniger/1989}). The use of ``wrong'' remainder estimates only
leads to a decrease of the efficiency of the transformation process
(compare for instance \cite[Theorem 13-12]{Weniger/1989}), depending on
the magnitude of $l$. With the help of a generalization of
Germain-Bonne's formal theory of convergence acceleration
\cite{Germain-Bonne/1973} it can be proved rigorously that the $u$ 
transformation accelerates linear convergence \cite[Theorems 12-10,
12-11, and 12-16]{Weniger/1989}.

Moreover, the $u$ transformation is also capable of summing effectively
many alternating divergent series. According to Smith and Ford
\cite{Smith/Ford/1979,Smith/Ford/1982} the $u$ transformation is among
the most versatile and powerful sequence transformations that are
currently known. This explains why Levin's $u$ transformation is used
internally in the computer algebra system Maple in the case of
convergence problems (see for example \cite[pp.\ 51 and
125]{Corless/2002} or \cite[p.\ 258]{Heck/2003}).

The remainder estimate (\ref{uRemEst}) can also be inserted into the
explicit expressions (\ref{GenWenTr}), (\ref{GenMinTr}), and
(\ref{GenCizTr}) for $\mathcal{S}_k^{(n)} (\beta, s_n, \omega_n)$,
$\mathcal{M}_k^{(n)} (\xi, s_n, \omega_n)$, and $\mathcal{C}_{k}^{(n)}
(\alpha, \beta, s_n, \omega_n)$, yielding the $u$-type variants
\cite[Eqs.\ (8.4-2) and (9.4-2)]{Weniger/1989}
\begin{align}
y_{k}^{(n)} (\beta, s_n) & \; = \;
\mathcal{S}_{k}^{(n)} (\beta, s_n, (\beta + n) \Delta s_{n-1})
\, , \qquad k, n \in \mathbb{N}_0 \, , 
\label{uWenTr} 
\\
Y_{k}^{(n)} (\xi, s_n) & \; = \;
\mathcal{M}_{k}^{(n)} (\xi, s_n, (-\xi-n) \Delta s_{n-1}) 
\, , \qquad k, n \in \mathbb{N}_0 \, , 
\label{uMinTr} 
\\
{\vphantom{\mathcal{C}}}_{u} \mathcal{C}_{k}^{(n)} 
(\alpha, \beta, s_n) & \; = \; 
\mathcal{C}_{k}^{(n)} (\alpha, \beta, s_n, (\beta + n) \Delta s_{n-1})
\, , \qquad k, n \in \mathbb{N}_0 \, .
\label{uCizTr}
\end{align}
In the case of the sequence transformation $\mathcal{G}_{k}^{(n)} (q_m,
s_n, \omega_n)$ introduced by \v{C}\'{\i}\v{z}ek, Zamastil, and
Sk\'{a}la \cite[Eq.\ (10)]{Cizek/Zamastil/Skala/2003} we choose the
$u$-type remainder estimates according to 
\begin{equation}
\omega_n \; = \; (n+q_0) \Delta s_{n-1} \; = \; (n+q_0) a_n \, ,
\label{GuRemEst}
\end{equation}
where $q_0 \ge 0$ is a suitable constant. Inserting this into
(\ref{GenCiZaSkaTr}) yields:
\begin{eqnarray}
\lefteqn{{\vphantom{\mathcal{G}}}_{u} \mathcal{G}_{k}^{(n)} (q_m, s_n) 
\; = \; 
\mathcal{G}_{k}^{(n)} (q_m, s_n, (n+q_0) \Delta s_{n-1})} \nonumber \\
& = \; \frac
{\displaystyle
\sum_{j=0}^{k} \, ( - 1)^{j} \, {\binom{k}{j}} \,
\prod_{m=1}^{k-1} \, \frac {(n+j+q_m)}{(n+k+q_m)} \,
\frac {s_{n+j}} {(n+j+q_0) \Delta s_{n+j-1}} }
{\displaystyle
\sum_{j=0}^{k} \, ( - 1)^{j} \, {\binom{k}{j}} \,
\prod_{m=1}^{k-1} \, \frac {(n+j+q_m)}{(n+k+q_m)} \,
\frac {1} {(n+j+q_0) \Delta s_{n+j-1}} } 
\, , \qquad k, n \in \mathbb{N}_0 \, . 
\label{uCiZaSkaTr}
\end{eqnarray}
As discussed in more details in Section \ref{Sec:RatApprox} (see the
discussion following (\ref{NumerEst1})), this $u$-type transformation
looses some important exactness properties if $q_0 > 0$ is essentially
arbitrary and does not satisfy $q_0 \in \{ q_1, \ldots, q_m \}$ because
then $\bigl[\prod_{m=1}^{k-1} (n+q_m) \bigr]/(n+q_0)$ is rational in $n$
and does not simplify to a polynomial of degree $k-2$ in $n$. Thus, an
obvious idea would be to choose $q_0 = q_1$. The other $u$-type
transformations (\ref{uLevTr}), (\ref{uWenTr}), (\ref{uMinTr}), and
(\ref{uCizTr}) satisfy this requirement.

The asymptotic expansion (\ref{TermLinConv}) indicates that in the case
linear convergence the term $a_n$ is a natural estimate for the
truncation error of the sequence (\ref{AsyLinLogModSeq}) with $\vert z
\vert < 1$. Thus, Levin \cite{Levin/1973} proposed for linearly 
convergent sequences and series the remainder estimate
\begin{equation}
\omega_n \; = \; \Delta s_{n-1} \; = \; a_n \, ,
\label{tRemEst}
\end{equation}
which yields Levin's $t$ transformation in the notation of \cite[Eq.\
(7.3-7)]{Weniger/1989}:
\begin{equation}
t_{k}^{(n)} (\beta, s_n) \; = \;
\mathcal{L}_{k}^{(n)} (\beta, s_n, \Delta s_{n-1}) 
\, , \qquad k, n \in \mathbb{N}_0 \, .
\label{tLevTr}
\end{equation}

The $t$ transformation is an effective accelerator for linear
convergence and in particular for alternating series
\cite{Smith/Ford/1979,Smith/Ford/1982}. With the help of a 
generalization of Germain-Bonne's formal theory of convergence
acceleration \cite{Germain-Bonne/1973} it can be proved rigorously that
Levin's $t$ transformation accelerates linear convergence \cite[Theorems
12-10, 12-11, and 12-16]{Weniger/1989}. It is also able to sum many
alternating divergent series. However, a comparison of
(\ref{AsyLinLogModSeq}) and (\ref{TermLogConv}) indicates that the $t$
transformation should fail to accelerate logarithmic convergence (for
more details, see \cite[Theorem 14-1]{Weniger/1989}).

The use of the remainder estimate (\ref{tRemEst}) in the explicit
expressions (\ref{GenWenTr}), (\ref{GenMinTr}), and (\ref{GenCizTr}) for
$\mathcal{S}_k^{(n)} (\beta, s_n, \omega_n)$, $\mathcal{M}_k^{(n)} (\xi,
s_n, \omega_n)$, and $\mathcal{C}_{k}^{(n)} (\alpha, \beta, s_n,
\omega_n)$ yields the $t$-type variants \cite[Eqs.\ (8.4-3) and
(9.4-3)]{Weniger/1989}
\begin{align}
\tau_{k}^{(n)} (\beta, s_n) & \; = \;
\mathcal{S}_{k}^{(n)} (\beta, s_n, \Delta s_{n-1})
\, , \qquad k, n \in \mathbb{N}_0 \, , 
\label{tWenTr} 
\\
T_{k}^{(n)} (\xi, s_n) & \; = \;
\mathcal{M}_{k}^{(n)} (\xi, s_n, \Delta s_{n-1}) 
\, , \qquad k, n \in \mathbb{N}_0 \, , 
\label{tMinTr} 
\\
{\vphantom{\mathcal{C}}}_{t} \mathcal{C}_{k}^{(n)} 
(\alpha, \beta, s_n) & \; = \; 
\mathcal{C}_{k}^{(n)} (\alpha, \beta, s_n, \Delta s_{n-1})
\, , \qquad k, n \in \mathbb{N}_0 \, .
\label{tCizTr}
\end{align}
In the case of the sequence transformation $\mathcal{G}_{k}^{(n)} (q_m,
s_n, \omega_n)$ we obtain in this way:
\begin{eqnarray}
\lefteqn{{\vphantom{\mathcal{G}}}_{t} \mathcal{G}_{k}^{(n)} (q_m, s_n) 
\; = \; 
\mathcal{G}_{k}^{(n)} (q_m, s_n, \Delta s_{n-1})} \nonumber \\
& = \; \frac
{\displaystyle
\sum_{j=0}^{k} \, ( - 1)^{j} \, {\binom{k}{j}} \,
\prod_{m=1}^{k-1} \, \frac {(n+j+q_m)}{(n+k+q_m)} \,
\frac {s_{n+j}} {\Delta s_{n+j-1}} }
{\displaystyle
\sum_{j=0}^{k} \, ( - 1)^{j} \, {\binom{k}{j}} \,
\prod_{m=1}^{k-1} \, \frac {(n+j+q_m)}{(n+k+q_m)} \,
\frac {1} {\Delta s_{n+j-1}} } 
\, , \qquad k, n \in \mathbb{N}_0 \, .
\label{tCiZaSkaTr}
\end{eqnarray}

Inspired by Aitken's $\Delta^2$ formula \cite{Aitken/1926}, Levin 
\cite{Levin/1973} introduced as a third simple remainder estimate
\begin{equation}
\omega_n \; = \; \frac{\Delta s_{n-1} \Delta s_n}
{\Delta s_{n-1} - \Delta s_n} \; = \; 
\frac{a_n a_{n+1}}{a_n - a_{n+1}} \, .
\label{vRemEst}
\end{equation}
The usefulness of this remainder estimate can be demonstrated by
applying it to the model sequence (\ref{AsyLinLogModSeq}). For $\zeta =
1$ and $\mathrm{Re} (\theta) < 0$ (logarithmic convergence) we obtain
\begin{eqnarray}
\lefteqn{\frac{a_n a_{n+1}}{a_n - a_{n+1}} \; = \; n^{\theta} \, 
\Biggl\{ - \, \frac{\theta \alpha_0}{\theta-1} \, - \, 
\frac{\theta \alpha_1}{(\theta-1) n}} \nonumber \\
& + \; {\displaystyle
\left[(\theta+1) \left\{ \frac{\theta  \alpha_0}{12} - 
\frac{(\theta-2) \alpha_2}{\theta-1} \right\} - 
\frac{{\alpha_1}^2}{\theta \alpha_0} \right] \,  
\frac{1}{(\theta-1)n^2} + \mathrm{O} \bigl( n^{-3} \bigr) \Biggr\}} \, ,
\qquad n \to \infty \, ,
\label{vModSeqLinConv}
\end{eqnarray}
and for $\vert z \vert < 1$ (linear convergence) we obtain
\begin{equation}
\frac{a_n a_{n+1}}{a_n - a_{n+1}} \; = \; z^{n+1} \, n^{\theta} \, 
\left\{ - \, \alpha_0 \, - \, \frac{\alpha_1}{n} \, - \,
\left[ \frac{z \theta \alpha_0}{(z-1)^2} + \alpha_2 \right] \,  
\frac{1}{n^2} + \mathrm{O} \bigl( n^{-3} \bigr) \right\} \, ,
\quad n \to \infty \, .
\label{vModSeqLogConv}
\end{equation}

It is a remarkable feature of the remainder estimate (\ref{vRemEst})
that it does not only reproduce the leading order of the model sequence
(\ref{AsyLinLogModSeq}), but both in the case of linear and logarithmic
convergence also the next one. Thus, the reminder estimate
(\ref{vRemEst}) is in the case of the model sequence
(\ref{AsyLinLogModSeq}) a remainder estimate of the type of
(\ref{ModAsyCondRemEst}) with $l = 2$, whereas Levin's other two
remainder estimates (\ref{uRemEst}) and (\ref{tRemEst}) only satisfy
(\ref{AsyCondRemEst}). Of course, it would be desirable to find other
remainder estimate which are also able to reproduce more than the
leading order asymptotics of the truncation error. Further research into
this direction should be of considerable interest.

The use of the remainder estimate (\ref{vRemEst}) in (\ref{GenLevTr})
yields Levin's $v$ transformation in the notation of \cite[Eq.\
(7.3-11)]{Weniger/1989}:
\begin{equation}
v_{k}^{(n)} (\beta, s_n) \; = \;
\mathcal{L}_{k}^{(n)} ( \beta, s_n,
\Delta s_{n-1} \Delta s_n / [\Delta s_{n-1} - \Delta s_n] )
\, , \qquad k, n \in \mathbb{N}_0 \, . 
\label{vLevTr}
\end{equation}

Levin's $v$ transformation is an effective accelerator for many linearly
and logarithmically convergent sequences and series. With the help of a
generalization of Germain-Bonne's formal theory of convergence
acceleration \cite{Germain-Bonne/1973} it can be proved rigorously that
the $v$ transformation accelerates linear convergence \cite[Theorems
12-10 and 12-11]{Weniger/1989}. The $v$ transformation is also able to
sum many alternating divergent series. According to Smith and Ford
\cite{Smith/Ford/1979,Smith/Ford/1982}, the $v$ transformation has
similar properties as the $u$ transformation, which means that it is
among the most versatile and powerful sequence transformations that are
currently known.

The use of the remainder estimate (\ref{vRemEst}) in the explicit
expressions (\ref{GenWenTr}), (\ref{GenMinTr}), and (\ref{GenCizTr}) for
$\mathcal{S}_k^{(n)} (\beta, s_n, \omega_n)$, $\mathcal{M}_k^{(n)} (\xi,
s_n, \omega_n)$, and $\mathcal{C}_{k}^{(n)} (\alpha, \beta, s_n,
\omega_n)$ yields the $v$-type variants \cite[Eqs.\ (8.4-5) and
(9.4-5)]{Weniger/1989}
\begin{align}
\varphi_{k}^{(n)} (\beta, s_n) & \; = \;
\mathcal{S}_{k}^{(n)} \bigl(\beta, s_n, 
\Delta s_{n-1} \Delta s_n / [\Delta s_{n-1} - \Delta s_n] \bigr)
\, , \qquad k, n \in \mathbb{N}_0 \, , 
\label{vWenTr} 
\\
\Phi_{k}^{(n)} (\xi, s_n) & \; = \;
\mathcal{M}_{k}^{(n)} \bigl(\xi, s_n, 
\Delta s_{n-1} \Delta s_n / [\Delta s_{n-1} - \Delta s_n] \bigr) 
\, , \qquad k, n \in \mathbb{N}_0 \, . 
\label{vMinTr} 
\\
{\vphantom{\mathcal{C}}}_{t} \mathcal{C}_{k}^{(n)} 
(\alpha, \beta, s_n) & \; = \; 
\mathcal{C}_{k}^{(n)} \bigl(\alpha, \beta, s_n, 
\Delta s_{n-1} \Delta s_n / [\Delta s_{n-1} - \Delta s_n] \bigr)
\, , \qquad k, n \in \mathbb{N}_0 \, .
\label{vCizTr}
\end{align}
In the case of the sequence transformation $\mathcal{G}_{k}^{(n)} (q_m,
s_n, \omega_n)$ we obtain
\begin{eqnarray}
\lefteqn{{\vphantom{\mathcal{G}}}_{v} \mathcal{G}_{k}^{(n)} (q_m, s_n) 
\; = \; \mathcal{G}_{k}^{(n)} \bigl(q_m, s_n, 
\Delta s_{n-1} \Delta s_n / [\Delta s_{n-1} - \Delta s_n]\bigr)} 
\nonumber \\
& = \; \frac
{\displaystyle
\sum_{j=0}^{k} \, ( - 1)^{j} \, {\binom{k}{j}} \,
\prod_{m=1}^{k-1} \, \frac {(n+j+q_m)}{(n+k+q_m)} \,
\frac{(\Delta s_{n+j-1} - \Delta s_{n+j}) s_{n+j}}
{\Delta s_{n+j-1} \Delta s_{n+j}} }
{\displaystyle
\sum_{j=0}^{k} \, ( - 1)^{j} \; {\binom{k}{j}} \,
\prod_{m=1}^{k-1} \, \frac {(n+j+q_m)}{(n+k+q_m)} \,
\frac {\Delta s_{n+j-1} - \Delta s_{n+j}}
{\Delta s_{n+j-1} \Delta s_{n+j} } }
\, , \quad k, n \in \mathbb{N}_0 \, .
\label{vCiZaSkaTr}
\end{eqnarray}

The best simple estimate for the truncation error of a strictly
alternating convergent series is the first term not included in the
partial sum \cite[p.\ 259]{Knopp/1964}. Moreover, the first term
neglected is also an estimate of the truncation error of a divergent
hypergeometric series ${}_2 F_0 (a, b, - z)$ with $a, b, z > 0$
\cite[Theorem 5.12-5]{Carlson/1977}. Accordingly, Smith and Ford 
\cite{Smith/Ford/1979} proposed the remainder estimate 
\begin{equation}
\omega_n \; = \; \Delta s_n \; = \; a_n \, ,
\label{dRemEst}
\end{equation}
This remainder estimate can also be motivated via the model sequence
(\ref{AsyLinLogModSeq}). For $z = 1$ and $\mathrm{Re}
(\theta) < 0$ (logarithmic convergence), we find
\begin{eqnarray}
a_{n+1} & = &  n^{\theta} \, \biggl\{ \frac{\theta \alpha_0}{n} +  
\frac{(\theta-1) [2 \alpha_1 + \theta \alpha_0]}{2 n^2} \nonumber \\
& & \quad + \, 
{\displaystyle \frac{(\theta-2) \bigl[ 6 \alpha_2 + (\theta-1)
\{ \theta \alpha_0 + 3 \alpha_1 \} \bigr]}{6 n^3}} 
\, + \, \mathrm{O} \bigl( n^{-3} \bigr) \biggr\} \, , 
\qquad n \to \infty \, ,
\label{NextTermLogConv}
\end{eqnarray}
and for $\vert z \vert < 1$ (linear convergence), we find.
\begin{eqnarray}
a_{n+1} & = & z^{n+1} \, n^{\theta} \, \biggl\{ (z-1) \alpha_0 + 
\frac{z \theta \alpha_0 + (z-1) \alpha_1}{n} \nonumber \\
&& \qquad + \, 
{\displaystyle \frac{z (\theta-1) \left[2 \alpha_1 + \theta
\alpha_0 \right] + 2 (z-1) \alpha_2}{2 n^2}}
\, + \, \mathrm{O} \bigl( n^{-3} \bigr) \biggr\} \, , 
\qquad n \to \infty \, .
\label{NextTermLinConv}
\end{eqnarray}

The use of the remainder estimate (\ref{dRemEst}) yields Levin's $d$
transformation in the notation of \cite[Eq.\ (7.3-9)]{Weniger/1989}:
\begin{equation}
d_{k}^{(n)} (\beta, s_n) \; = \;
\mathcal{L}_{k}^{(n)} (\beta, s_n, \Delta s_n) 
\, , \qquad k, n \in \mathbb{N}_0 \, .
\label{dLevTr}
\end{equation}

If we compare (\ref{NextTermLogConv}) with (\ref{TermLogConv}) and
(\ref{NextTermLinConv}) with (\ref{TermLinConv}), we see that Levin's
$d$ transformation should have similar properties as Levin's $t$
transformation. This is confirmed by experience: The $d$ transformation
is a powerful accelerator for linear convergence and in particular for
alternating series, and is also able to sum many alternating divergent
series, but fails to accelerate logarithmic convergence. With the help
of a generalization of Germain-Bonne's formal theory of convergence
acceleration \cite{Germain-Bonne/1973} it can be proved rigorously that
the $d$ transformation accelerates linear convergence \cite[Theorems
12-10, 12-11, and 12-16]{Weniger/1989}.

The use of the remainder estimate (\ref{dRemEst}) in the explicit
expressions (\ref{GenWenTr}), (\ref{GenMinTr}), and (\ref{GenCizTr}) for
$\mathcal{S}_k^{(n)} (\beta, s_n, \omega_n)$, $\mathcal{M}_k^{(n)} (\xi,
s_n, \omega_n)$, and $\mathcal{C}_{k}^{(n)} (\alpha, \beta, s_n,
\omega_n)$ yields the $d$-type variants \cite[Eqs.\ (8.4-4) and
(9.4-4)]{Weniger/1989}
\begin{align}
\delta_{k}^{(n)} (\beta, s_n) & \; = \;
\mathcal{S}_{k}^{(n)} (\beta, s_n, \Delta s_n)
\, , \qquad k, n \in \mathbb{N}_0 \, , 
\label{dWenTr} 
\\
\Delta_{k}^{(n)} (\xi, s_n) & \; = \;
\mathcal{M}_{k}^{(n)} (\xi, s_n, \Delta s_n) 
\, , \qquad k, n \in \mathbb{N}_0 \, , 
\label{dMinTr} 
\\
{\vphantom{\mathcal{C}}}_{d} \mathcal{C}_{k}^{(n)} 
(\alpha, \beta, s_n) & \; = \; 
\mathcal{C}_{k}^{(n)} (\alpha, \beta, s_n, \Delta s_n)
\, , \qquad k, n \in \mathbb{N}_0 \, .
\label{dCizTr}
\end{align}

As already mentioned in Section \ref{Sec:Intro}, the delta
transformation (\ref{dWenTr}) was found to be particularly powerful in
the case of factorially and hyperfactorially divergent alternating power
series as they for instance occur in the perturbation expansions of
quantum physics
\cite{Borghi/Santarsiero/2003,Cizek/Vinette/Weniger/1991,
Cizek/Vinette/Weniger/1993,Cizek/Zamastil/Skala/2003,%
Jentschura/Becher/Weniger/Soff/2000,Jentschura/Weniger/Soff/2000,
Weniger/1989,Weniger/1990,Weniger/1992,Weniger/1994a,Weniger/1994b,
Weniger/1996a,Weniger/1996b,Weniger/1996c,Weniger/1996e,Weniger/1997,%
Weniger/2001,Weniger/Cizek/Vinette/1991,Weniger/Cizek/Vinette/1993} or
in asymptotic expansions for special functions
\cite{Weniger/1990,Weniger/1994b,Weniger/1996d,Weniger/Cizek/1990}).

In the case of the sequence transformation $\mathcal{G}_{k}^{(n)} (q_m,
s_n, \omega_n)$ we obtain
\begin{eqnarray}
\lefteqn{{\vphantom{\mathcal{G}}}_{d} \mathcal{G}_{k}^{(n)} (q_m, s_n) 
\; = \; 
\mathcal{G}_{k}^{(n)} (q_m, s_n, \Delta s_n)} \nonumber \\
& = \; \frac
{\displaystyle
\sum_{j=0}^{k} \, ( - 1)^{j} \; {\binom{k}{j}} \,
\prod_{m=1}^{k-1} \, \frac {(n+j+q_m)}{(n+k+q_m)} \,
\frac {s_{n+j}} {\Delta s_{n+j}} }
{\displaystyle
\sum_{j=0}^{k} \, ( - 1)^{j} \, {\binom{k}{j}} \,
\prod_{m=1}^{k-1} \, \frac {(n+j+q_m)}{(n+k+q_m)} \,
\frac {1} {\Delta s_{n+j}} } 
\, , \qquad k, n \in \mathbb{N}_0 \, . 
\label{dCiZaSkaTr}
\end{eqnarray}

In practical applications it happens relatively often that asymptotic
expressions $a_{n}^{(\infty)}$ for the terms $a_n$ of a series are known
that reproduce the leading order of $a_n$ as $n \to \infty$. For
example, the coefficients $c_n$ of many divergent perturbation
expansions, which are power series in some coupling constant $g$,
satisfy (see for example \cite[Table 1]{Fischer/1997})
\begin{equation}
c_n \; = \; (-1)^n \, \Gamma (an + b) \, R^n \, \bigl[ C \, + \,
\mathrm{O} (1/n) \bigr] \, , \qquad n \to \infty \, ,
% \label{}
\end{equation}
where $a$, $b$, $C$, and $R$ are known constants. 

The remainder estimate (\ref{vRemEst}), which leads to Levin's $v$
transformation, is in some sense exceptional since reproduces not only
the leading order of the truncation errors of the model sequence
(\ref{AsyLinLogModSeq}) as $n \to \infty$, but also the next
one. Normally, we can only expect that the simple remainder estimates
(\ref{uRemEst}), (\ref{tRemEst}), (\ref{vRemEst}), and (\ref{dRemEst})
reproduce the leading order of the remainder $s_n - s$. However, the
leading order of the truncation error is also reproduced if we use in
(\ref{uRemEst}), (\ref{tRemEst}), (\ref{vRemEst}), and (\ref{dRemEst})
not $a_n$ and $a_{n+1}$ but their limiting expressions $a^{(\infty)}_n$
and $a^{(\infty)}_{n+1}$. Whether this improves the transformation
results or not, depends on the problem under consideration and cannot be
decided by simple considerations. Nevertheless, it may well be worth a
try. Ideas of that kind were discussed in more details in
\cite{Homeier/Weniger/1995,Weniger/1994a} and also by
\v{C}\'{\i}\v{z}ek, Zamastil, and Sk\'{a}la 
\cite{Cizek/Zamastil/Skala/2003}. 

The $t$-type transformations (\ref{tLevTr}), (\ref{tWenTr}),
(\ref{tMinTr}), (\ref{tCizTr}), and (\ref{tCiZaSkaTr}) use the last term
$a_n$ of the partial sum $s_n = \sum_{k=0}^{n} a_k$ as an estimate for
the truncation error $r_n = - \sum_{k=n+1}^{\infty} a_k$, whereas the
analogous $d$-type transformations (\ref{dLevTr}), (\ref{dWenTr}),
(\ref{dMinTr}), (\ref{dCizTr}), and (\ref{dCiZaSkaTr}) use the first
term $a_{n+1}$ not included in the partial sum as the remainder
estimate. Thus, it looks that $t$-type transformation utilize the
available information in some sense more effectively than $d$-type
transformation since they use $a_{n+1}$ also for the construction of the
next partial sum $s_{n+1}$. This is, however, a superficial judgment and
it can be shown easily that $t$-type transformations are actually
$d$-type transformations is disguise.

All $t$-type transformations of this article can be expressed
as follows:
\begin{align}
{\vphantom{\mathcal{T}}}_{t} T_{k}^{(n)} (s_n, \Delta s_{n-1}) & \; = \;
\frac
{\displaystyle \Delta^k \, [P_{k-1} (n) \, s_n/\Delta s_{n-1}]}
{\displaystyle \Delta^k \, [P_{k-1} (n)/\Delta s_{n-1}]}
\label{Gen_t-TypeTrDiffOpRep} \\
& \; = \; 
\frac
{\displaystyle
\sum_{j=0}^{k} \; ( - 1)^{j} \; {\binom{k}{j}} \;
P_{k-1} (n+j) \, \frac {s_{n+j}} {\Delta s_{n+j-1}} }
{\displaystyle
\sum_{j=0}^{k} \; ( - 1)^{j} \; {\binom{k}{j}} \;
P_{k-1} (n+j) \, \frac {1} {\Delta s_{n+j-1}} } \, ,
\qquad k \in \mathbb{N} \, , \quad n \in \mathbb{N}_0 \, .
\label{Gen_t-TypeTr}
\end{align}
As usual, $P_{k-1} (n)$ is a polynomial of degree $k-1$ in $n$.

Let us now use the relationship $s_n = s_{n-1} + \Delta s_{n-1}$ in the
numerator on the right-hand side of (\ref{Gen_t-TypeTrDiffOpRep}). In
view of the linearity of $\Delta^k$ we then obtain:
\begin{equation}
\Delta^k \, \frac{P_{k-1} (n) s_n}{\Delta s_{n-1}}
\; = \;
\Delta^k \, \left\{ P_{k-1} (n) \, \left[ \frac{s_{n-1}}{\Delta s_{n-1}} 
+ \frac{\Delta s_{n-1}}{\Delta s_{n-1}} \right] \right\} \; = \;
\Delta^k \, \frac{P_{k-1} (n) s_{n-1}}{\Delta s_{n-1}} \, .
% \label{}
\end{equation}
Inserting this into (\ref{Gen_t-TypeTrDiffOpRep}) and
(\ref{Gen_t-TypeTr}) shows that a $t$-type transformation is actually a
$d$-type transformation with a transformed polynomial $\mathcal{P}_{k-1}
(n-1) = P_{k-1} (n)$:
\begin{align}
{\vphantom{\mathcal{T}}}_{t} T_{k}^{(n)} (s_n, \Delta s_{n-1}) & 
\; = \; \frac
{\displaystyle \Delta^k \, [\mathcal{P}_{k-1} (n-1) \, s_{n-1}/
\Delta s_{n-1}]}
{\displaystyle \Delta^k \, [\mathcal{P}_{k-1} (n-1)/\Delta s_{n-1}]}
\label{Mod_t-TypeTrDiffOpRep} \\
& \; = \; 
\frac
{\displaystyle
\sum_{j=0}^{k} \; ( - 1)^{j} \; {\binom{k}{j}} \;
\mathcal{P}_{k-1} (n+j-1) \, \frac {s_{n+j-1}} {\Delta s_{n+j-1}} }
{\displaystyle
\sum_{j=0}^{k} \; ( - 1)^{j} \; {\binom{k}{j}} \;
\mathcal{P}_{k-1} (n+j-1) \, \frac {1} {\Delta s_{n+j-1}} } \, .
\label{Mod_t-TypeTr}
\end{align}

Thus, the $t$-type transformations defined in (\ref{tLevTr}),
(\ref{tWenTr}), (\ref{tMinTr}), (\ref{tCizTr}), and (\ref{tCiZaSkaTr})
can be expressed by the corresponding $d$-type transformations defined
in (\ref{dLevTr}), (\ref{dWenTr}), (\ref{dMinTr}), (\ref{dCizTr}), and
(\ref{dCiZaSkaTr}) according to
\begin{align}
t_{k}^{(n)} (\beta, s_n) & \; = \;
\mathcal{L}_{k}^{(n-1)} (\beta+1, s_{n-1}, \Delta s_{n-1}) \; = \;
d_{k}^{(n-1)} (\beta+1, s_{n-1})
\, , \label{Mod_tLevTr} 
\\
\tau_{k}^{(n)} (\beta, s_n) & \; = \; 
\mathcal{S}_{k}^{(n-1)} (\beta+1, s_{n-1}, \Delta s_{n-1}) \; = \;
\delta_{k}^{(n-1)} (\beta+1, s_{n-1})
\, , \label{Mod_tWenTr} 
\\
T_{k}^{(n)} (\xi, s_n) & \; = \; 
\mathcal{M}_{k}^{(n-1)} (\xi+1, s_{n-1}, \Delta s_{n-1}) \; = \;
\Delta_{k}^{(n-1)} (\xi+1, s_{n-1})
\, , \label{Mod_tMinTr} 
\\
{\vphantom{\mathcal{C}}}_{t} \mathcal{C}_{k}^{(n)} 
(\alpha, \beta, s_n) & \; = \; 
\mathcal{C}_{k}^{(n-1)} (\alpha, \beta+1, s_{n-1}, \Delta s_{n-1}) 
\; = \; {\vphantom{\mathcal{C}}}_{d} \mathcal{C}_{k}^{(n)} 
(\alpha, \beta+1, s_{n-1})
\, , \label{Mod_tCizTr}
\\
{\vphantom{\mathcal{G}}}_{t} \mathcal{G}_{k}^{(n)} (q_m, s_n) 
& \; = \;   
\mathcal{G}_{k}^{(n-1)} (q_m+1, s_{n-1}, \Delta s_{n-1} \; = \; 
{\vphantom{\mathcal{G}}}_{d} \mathcal{G}_{k}^{(n-1)} (q_m+1, s_{n-1})
\, . 
\label{Mod_tCiZaSkaTr} 
\end{align}

In these expressions, the case $n = 0$ deserves special
consideration. Since we tacitly assume $s_{-m} = 0$ with $m \in
\mathbb{N}$, the term with $j=0$ in the numerator sum of
(\ref{Mod_t-TypeTr}) vanishes for $n = 0$. This can also be proved
directly from the numerator sum in (\ref{Gen_t-TypeTr}). If we write
there $\Delta s_{n+j-1} = a_{n+j}$ and $s_{n+j} = \sum_{\nu=0}^{n+j}
a_{\nu}$, we obtain for $n = 0$:
\begin{align}
\sum_{j=0}^{k} \; ( - 1)^{j} \; {\binom{k}{j}} \; P_{k-1} (j) \, 
\sum_{\nu=0}^{j} \frac {a_{\nu}}{a_j} 
& \; = \;
\sum_{j=0}^{k} \; ( - 1)^{j} \; {\binom{k}{j}} \; P_{k-1} (j) \, 
\biggl[ \, \sum_{\nu=0}^{j-1} \frac {a_{\nu}}{a_j}  \, + \, 
\frac{a_j}{a_j} \biggr] 
\\
& \; = \;
\sum_{j=1}^{k} \; ( - 1)^{j} \; {\binom{k}{j}} \; P_{k-1} (j) \, 
\sum_{\nu=0}^{j-1} \frac {a_{\nu}}{a_j} \, .
\end{align}
Here, we made use of the fact that $\sum_{\nu=0}^{j-1} a_{\nu}$ is for
$j=0$ an empty sum which is zero.  Thus, for $n = 0$ the general
$t$-type transformation (\ref{Gen_t-TypeTr}) can be expressed as
follows:
\begin{align}
{\vphantom{\mathcal{T}}}_{t} T_{k}^{(0)} (s_0, \Delta s_{-1}) & \; = \;
\frac
{\displaystyle
\sum_{j=1}^{k} \; ( - 1)^{j} \; {\binom{k}{j}} \;
P_{k-1} (j) \, \frac {s_{j-1}} {\Delta s_{j-1}} }
{\displaystyle
\sum_{j=0}^{k} \; ( - 1)^{j} \; {\binom{k}{j}} \;
P_{k-1} (j) \, \frac {1} {\Delta s_{j-1}} } 
% \label{}
\\
& \; = \; 
\frac
{\displaystyle
\sum_{j=1}^{k} \; ( - 1)^{j} \; {\binom{k}{j}} \;
\mathcal{P}_{k-1} (j-1) \, \frac {s_{j-1}} {\Delta s_{j-1}} }
{\displaystyle
\sum_{j=0}^{k} \; ( - 1)^{j} \; {\binom{k}{j}} \;
\mathcal{P}_{k-1} (j-1) \, \frac {1} {\Delta s_{j-1}} } 
\, , \qquad k \in \mathbb{N} \, .
% \label{}
\end{align}

The fact that $t$-type transformations are actually $d$-type
transformations in disguise can also be deduced from the recursive
scheme (\ref{RecSchemeCiZaSkaTr}) which contains all the other recursive
schemes of Section \ref{Sec:RecForm} as special cases. If we choose the
initial conditions in (\ref{RecSchemeCiZaSkaTr_a}) according to $u_n =
s_n/\Delta s_{n-1}$, then (\ref{RecSchemeCiZaSkaTr_b}) implies
$G_{1}^{(n)} = \Delta [s_n/\Delta s_{n-1}] =
\Delta [s_{n-1}/\Delta s_{n-1}]$. Thus, (\ref{RecSchemeCiZaSkaTr_c}) 
with $k \ge 1$ yields the same results as if we had started the
recursion with the $d$-type initial conditions $u_n = s_{n-1}/\Delta
s_{n-1}$.

Numerical cancellation increases the risk of loosing accuracy. Thus, it
is probably wiser not to use the $t$-type initial conditions $u_n =
s_n/\Delta s_{n-1}$ in the recursive scheme (\ref{RecSchemeCiZaSkaTr})
or in any of its special cases, but instead the $d$-type initial
conditions $u_n = s_{n-1}/\Delta s_{n-1}$.

\typeout{==> Richardson-Type Transformations}
\section{Richardson-Type Transformations}
\label{Sec:RichTypeTrans}

Some of the most effective accelerators for logarithmically convergent
sequences and series ($\rho = 1$ in (\ref{DefLinLogConv})), which abound
in scientific applications and which constitute formidable computational
problems, can be derived with the help of interpolation theory. Thus,
the existence of a function $S$ of a continuous variable is postulated
which coincides on a set of discrete arguments $\{ x_n
\}_{n=0}^{\infty}$ with the elements of the sequence $\{ s_n
\}_{n=0}^{\infty}$ to be transformed:
\begin{equation}
S (x_n) \; = \; s_n \, , \qquad n \in \mathbb{N}_0 \, .
% \label{}
\end{equation}
This ansatz reduces the convergence acceleration problem to an
extrapolation problem. If a finite string $s_n$, $s_{n+1}$, $\ldots$,
$s_{n+k}$ of $k+1$ sequence elements is known, one can construct an
approximation $S_k (x)$ to $S (x)$ which satisfies the $k+1$
interpolation conditions $S_k (x_{n+j}) = s_{n+j}$ with $0 \le j \le k$.
Next, the value of $S_k (x)$ has to be determined for $x \to
x_{\infty}$. If this can be done, $S_k (x_{\infty})$ should provide a
better approximation to the limit $s = s_{\infty}$ of the sequence $\{
s_n \}_{n=0}^{\infty}$ than the last sequence element $s_{n+k}$ used for
its construction.

The most important interpolating functions are either polynomials or
rational functions which lead to different convergence algorithm (see
for example \cite[Section 6]{Weniger/1989}). Here, only polynomial
interpolation will be considered. Thus, it is assumed that the $k$-th
order approximant $S_k (x)$ is a polynomial of degree $k$ in $x$,
\begin{equation}
S_k (x) \; = \, \gamma_0 \, + \, \gamma_1 x \, + \, \cdots
\, + \, \gamma_k x^k \, , \qquad k \in \mathbb{N} \, ,
\label{InterpolPol}
\end{equation}
or equivalently that the model sequence for the Richardson extrapolation
scheme \cite{Richardson/1927}, whose construction will be sketched
below, is a polynomial of degree $k$ in the interpolation points $x_n$,
\begin{equation}
s_n \; = \, s \, + \, \sum_{j=0}^{k-1} \, c_j {x_n}^{j+1} \, ,
\qquad k \in \mathbb{N} \, , \quad n \in \mathbb{N}_0 \, .
\label{ModSeqRichAlg}
\end{equation}

For polynomials, the most natural extrapolation point is $x_{\infty} =
0$.  Accordingly, we assume that the interpolation points $x_n$ satisfy
the conditions
\begin{subequations}
\label{x_n2zero}
\begin{eqnarray}
& x_0 > x_1 > \cdots > x_m > x_{m+1} > \cdots > 0 \, ,
% \label{}
\\
& {\displaystyle \lim_{n \to \infty} \; x_n \; = \; 0} \, .
% \label{}
\end{eqnarray}
\end{subequations}
The choice $x_{\infty}=0$ implies that the approximation to the limit $s
= s_{\infty}$ in (\ref{ModSeqRichAlg}) is to be identified with the
constant term $\gamma_0$ of the polynomial (\ref{InterpolPol}).

Several different methods for the construction of interpolating
polynomials $S_k (x)$ are known (see for example \cite[Chapter
III]{Cuyt/Wuytack/1987}). Since we are only interested in the constant
term $\gamma_0$ of an interpolating polynomial $S_k (x)$ and since in
most applications it is desirable to compute simultaneously a whole
string of approximants $S_0 (0), S_1 (0), S_2 (0), \ldots$ with
increasing polynomial degree, the most economical choice is Neville's
scheme \cite{Neville/1934} for the recursive computation of
interpolating polynomials. If we set $x=0$ in Neville's scheme, we
obtain the following recursive scheme (see for example \cite[p.\
73]{Brezinski/RedivoZaglia/1991a} or \cite[Eq.\ (6.1-5)]{Weniger/1989}):
\begin{subequations}
\label{RichAl}
\begin{align}
\mathcal{N}_0^{(n)} (s_n, x_n) & \; = \; s_n \, ,
\quad n \in \mathbb{N}_0 \, ,
% \label{}
\\
\mathcal{N}_{k+1}^{(n)} (s_n, x_n) & \; = \;
\frac
{x_n \mathcal{N}_{k}^{(n+1)} (s_{n+1}, x_{n+1}) \, - \,
x_{n+k+1} \mathcal{N}_{k}^{(n)}  (s_n, x_n)}
{x_n \, - \, x_{n+k+1}} \; ,
\quad k,n \in \mathbb{N}_0 \, . \; \; \;
% \label{}
\end{align}
\end{subequations}
In the literature on convergence acceleration, this variant of Neville's
scheme is called Richardson extrapolation \cite{Richardson/1927}. In
\cite[Section 7.4]{Weniger/1989} it was shown that this recursive scheme
can also be derived with the help of the of the annihilation operator
approach described in Section \ref{Sec:AnnihilOp} by using divided
differences as annihilation operators.

In most applications, Richardson extrapolation is used in combination
with the interpolation points $x_n = 1/(n+\beta)$ with $\beta >
0$. Then, the model sequence (\ref{ModSeqRichAlg}) assumes the following
form:
\begin{equation}
s_n \; = \; s \, + \, \frac{1}{\beta+n} \, 
\sum_{j=0}^{k-1} \, \frac{c_j}{(\beta+n)^j} \, ,
\qquad k \in \mathbb{N} \, , \quad n \in \mathbb{N}_0 \, .
% \label{}
\end{equation}
This model sequence can be obtained from the model sequence
(\ref{LevTrModSeq}) for Levin's sequence transformation by setting
$\omega_n = 1/(\beta+n)$. Consequently, $\mathcal{N}_k^{(n)}$ with $x_n
= 1/(\beta+n)$ is a special Levin transformation and can be expressed as
the ratio of two finite sums according to (\ref{GenLevTr}). Since,
however, the denominator of the ratio (\ref{GenLevTr}) can for $\omega_n
= 1/(\beta+n)$ be expressed in closed form, $\mathcal{N}_k^{(n)}$
possesses an even simpler closed form expression as a finite sum (see
for example
\cite[Lemma 2.1, p.\ 313]{Marchuk/Shaidurov/1983} or
\cite[Eq.\ (7.3-20)]{Weniger/1989}):
\begin{align}
\Lambda_k^{(n)} (\beta, s_n) & \; = \; 
\mathcal{N}_k^{(n)} \bigl( s_n, 1/(\beta+n) \bigr) \; = \;
\mathcal{L}_{k}^{n} \bigl(\beta, s_n, 1/(\beta+n) \bigr) 
\nonumber \\
& \; = \; (-1)^k \, \sum_{j=0}^k \,
(-1)^j \, \frac {(\beta+n+j)^k} {j! \, (k-j)!} \, s_{n+j} \, ,
\quad k, n \in \mathbb{N}_0 \, .
\label{RichAlLev}
\end{align}
Moreover, the recursive scheme (\ref{RichAl}) assumes the following form
\cite[Eq.\ (7.3-21)]{Weniger/1989}:
\begin{subequations}
\label{RichAlLevRec}
\begin{align}
\Lambda_0^{(n)} (\beta, s_n) & \; = \; s_n \, ,
\qquad n \in \mathbb{N}_0 \, ,
% \label{}
\\
\Lambda_{k+1}^{(n)} (\beta, s_n) & \; = \;
\Lambda_k^{(n+1)} (\beta, s_{n+1}) \, + \,
\frac {\beta+n} {k+1} \, \Delta \Lambda_k^{(n)} (\beta, s_n)
\, , \qquad k, n \in \mathbb{N}_0 \, .
% \label{}
\end{align}  
\end{subequations}
In the case of doubly indexed quantities like $\Lambda_k^{(n)}$ it is
always assumed that $\Delta$ only acts on the superscript $n$ but not on
the subscript $k$, i.e., $\Delta \Lambda_k^{(n)} = \Lambda_k^{(n+1)} -
\Lambda_k^{(n)}$.

It is also possible to construct in the case of the sequence
transformation (\ref{GenCiZaSkaTr}) introduced by \v{C}\'{\i}\v{z}ek,
Zamastil, and Sk\'{a}la \cite[Eq.\ (10)]{Cizek/Zamastil/Skala/2003} a
Richardson-type variant. For that purpose, we set in the model sequence
(\ref{CiZaSkaTrModSeq}) $\omega_n = 1/(n+q_0)$, where $q_0$ is a
suitable constant, yielding
\begin{equation}
s_n \; = \; s \, + \,
\sum_{j=0}^{k-1} \, \frac{c_j}{\prod_{m=0}^{j} (n+q_m)} \, ,
\qquad k, n \in \mathbb{N}_0 \, .
\label{linCiZaSkaTrModSeq}
\end{equation}
Thus, $\Delta^k \prod_{m=0}^{k-1} (n+q_m)$ is the annihilation operator
for the remainder of this model sequence, yielding the following
Richardson-type variant of the sequence transformation introduced by
\v{C}\'{\i}\v{z}ek, Zamastil, and Sk\'{a}la \cite[Eq.\
(10)]{Cizek/Zamastil/Skala/2003}:
\begin{equation}
{\vphantom{\mathcal{G}}}_{R} \mathcal{G}_{k}^{(n)} (q_m, s_n) 
\; = \; \mathcal{G}_{k}^{(n)} (q_m, s_n, n+q_0) \; = \;
\frac{\Delta^k \bigl[ \prod_{m=0}^{k-1} (n+q_m) s_n \bigr]}
{\Delta^k \prod_{m=0}^{k-1} (n+q_m)} \, ,
\qquad k, n \in \mathbb{N}_0 \, .
\label{linCiZaSkaTrDiffOp}
\end{equation}
Of course, this transformation can be expressed as the ratio of two
finite sums according to (\ref{GenCiZaSkaTr}). However, the denominator
of (\ref{linCiZaSkaTrDiffOp}) can be expressed in closed form. First, we
observe that the products in the difference operator representation
(\ref{linCiZaSkaTrDiffOp}) are polynomials of degree $k$ in $n$,
satisfying
\begin{equation}
\prod_{m=0}^{k-1} (n+q_m) \; = \; n^k \, + \, 
(q_0 + q_1 + \ldots + q_{k-1}) n^{k-1} \, + \, \ldots \, .
% \label{}
\end{equation}
Next, we use the well known relationship $\Delta^k n^k = k!$, which can
for instance be derived by iterating the commutator relationship
(\ref{Comm1}), and take into account that all polynomials of degree $0$,
$1$, $\ldots$, $k-1$ in $n$ are annihilated by $\Delta^k$. Thus,
\begin{equation}
\Delta^k \prod_{m=0}^{k-1} (n+q_m) \; = \; k! \, .
\label{ClosedExprDenom}
\end{equation}
With the help of (\ref{DeltaPk}) we then obtain from
(\ref{linCiZaSkaTrDiffOp}): 
\begin{align}
{\vphantom{\mathcal{G}}}_{R} \mathcal{G}_{k}^{(n)} (q_m, s_n) 
& \; = \; \mathcal{G}_{k}^{(n)} (q_m, s_n, 1/(n+q_0)) \nonumber \\
& \; = \; (-1)^k \, \sum_{j=0}^{k-1} \, (-1)^j \, 
\frac{\prod_{m=0}^{k-1} (n+j+q_m)}{j! (k-j)!} \, s_{n+j} \, ,
\qquad k, n \in \mathbb{N}_0 \, .
\label{linCiZaSkaTr}
\end{align}
If we set here $q_m = \beta$, we obtain the sequence transformation
(\ref{RichAlLev}) according to
\begin{equation}
{\vphantom{\mathcal{G}}}_{R} \mathcal{G}_{k}^{(n)} (\beta, s_n) \; = \;
\mathcal{G}_{k}^{(n)} (\beta, s_n, 1/(\beta+n)) \; = \;
\Lambda_k^{(n)} (\beta, s_n) \, ,
\qquad k, n \in \mathbb{N}_0 \, .
% \label{}
\end{equation}

We can derive a recursive scheme for ${\vphantom{\mathcal{G}}}_{R}
\mathcal{G}_{k}^{(n)} (q_m, s_n)$ by means of the techniques described 
in Section \ref{Sec:RecForm}. For that purpose, we express the numerator
of the ratio on the right-hand side of (\ref{linCiZaSkaTrDiffOp}) as
follows:
\begin{align}
Q_{k}^{(n)} & \; = \; Q_{k}^{(n)} (q_m, s_n) \; = \;
\Delta^k Y_{k}^{(n)} \, , \qquad k, n \in \mathbb{N}_0 \, , 
\label{DefQkn} \\
Y_{k}^{(n)} & \; = \; Y_{k}^{(n)} (q_m, s_n) \; = \;
\prod_{m=0}^{k-1} (n+q_m) \, s_n \, , \qquad k, n \in \mathbb{N}_0 \, .
\label{DefYkn}
\end{align}

The quantities $Y_{k}^{(n)}$ satisfy for $k \ge 1$ the two-term
recursion
\begin{equation}
Y_{k}^{(n)} \; = \; (n+q_{k-1}) \, Y_{k-1}^{(n)} \, .
\label{RecYkn}
\end{equation}
Next, we combine (\ref{DefQkn}) - (\ref{RecYkn}) with the commutator
relationship (\ref{comm4}), yielding
\begin{align}
Q_{k}^{(n)} & \; = \; \Delta^k \, (n+q_{k-1}) \, Y_{k-1}^{(n)} \\
& \; = \; \bigl[ (n+k+q_{k-1}) \, E \, - \, (n+q_{k-1}) \bigr] 
\, \Delta^{k-1}  \, Y_{k-1}^{(n)} \\
& \; = \; (n+k+q_{k-1}) \, Q_{k-1}^{(n+1)} \, - \, 
(n+q_{k-1}) \, Q_{k-1}^{(n)} \, , 
\qquad k \in \mathbb{N} \, , \quad n \in \mathbb{N}_0 \, .
\label{RecQkn}
\end{align}
Now, we only have to divide the recurrence formula (\ref{RecQkn}) for
the numerator of (\ref{linCiZaSkaTrDiffOp}) by the denominator according
to (\ref{ClosedExprDenom}) to obtain the recursive scheme
\begin{subequations}
\label{linCiZaSkaTrRecScheme}
\begin{align}
{\vphantom{\mathcal{G}}}_{R} \mathcal{G}_{k}^{(0)} (q_m, s_n) 
& \; = \; s_n \, , \qquad n \in \mathbb{N}_0 \, ,
% \label{}
\\
{\vphantom{\mathcal{G}}}_{R} \mathcal{G}_{k+1}^{(n)} (q_m, s_n) & \; = \;
{\vphantom{\mathcal{G}}}_{R} \mathcal{G}_{k}^{(n+1)} (q_m, s_{n+1}) \, + \,
\frac {n + q_{k}} {k+1} \, \Delta 
{\vphantom{\mathcal{G}}}_{R} \mathcal{G}_{k}^{(n)} (q_m, s_n)
\, , \qquad k, n \in \mathbb{N}_0 \, .
% \label{}
\end{align}
\end{subequations}
If we choose in (\ref{linCiZaSkaTrRecScheme}) $q_m = \beta$, we obtain
the recursive scheme (\ref{RichAlLevRec}) for $\Lambda_k^{(n)} (\beta,
s_n)$.

By specializing the parameters $q_m$ in (\ref{linCiZaSkaTr}) and
(\ref{linCiZaSkaTrRecScheme}), other Richardson-type transformations and
their recursive schemes can be obtained. If we choose $q_m =
\chi + m$, we obtain the Richardson-type variant of the sequence
transformation (\ref{GenWenTr}) \cite[Eq.\ (8.4-11)]{Weniger/1989}.
\begin{align}
\mathcal{F}_k^{(n)} (\chi, s_n) & \; = \; 
{\vphantom{\mathcal{G}}}_{R} \mathcal{G}_{k}^{(n)} (\chi+m, s_n) \; = \;
\mathcal{S}_k^{(n)} \bigl( \chi+1, s_n, 1/(\chi+n) \bigr) \nonumber \\
& \; = \; 
(-1)^k \, {\displaystyle \sum_{j=0}^k \,
(-1)^j \, \frac {(\chi+n+j)_k} {j! (k-j)!} \, s_{n+j}}
\, , \qquad k, n \in \mathbb{N}_0 \, ,
\label{linWenTr}
\end{align}
and its recursive scheme \cite[Eq.\ (8.4-12)]{Weniger/1989}
\begin{subequations}
\label{linWenTrRecScheme}
\begin{align}
\mathcal{F}_0^{(n)} (\chi, s_n) & \; = \; s_n \, ,
\qquad n \in \mathbb{N}_0 \, ,
% \label{}
\\
\mathcal{F}_{k+1}^{(n)} (\chi, s_n) & \; = \;
\mathcal{F}_k^{(n+1)} (\chi, s_{n+1}) \, + \,
\frac {\chi+n+k} {k+1} \, \Delta \mathcal{F}_k^{(n)} (\chi, s_n)
\, , \qquad k, n \in \mathbb{N}_0 \, .
% \label{}
\end{align}  
\end{subequations}

If we choose in (\ref{linCiZaSkaTr}) and (\ref{linCiZaSkaTrRecScheme})
$q_m = \zeta - m$, we obtain the Richardson-type variant of the sequence
transformation (\ref{GenMinTr}) \cite[Eq.\ (9.4-11)]{Weniger/1989},
\begin{align}
\mathcal{P}_k^{(n)} (\zeta, s_n) & \; = \; 
{\vphantom{\mathcal{G}}}_{R} \mathcal{G}_{k}^{(n)} (\zeta-m, s_n) \; = \;
\mathcal{M}_k^{(n)} \bigl( \zeta-1, s_n, - 1/(\zeta+n) \bigr) \nonumber \\
& \; = \; 
{\displaystyle \sum_{j=0}^k \,
(-1)^j \, \frac {(-\zeta-n-j)_k} {j! (k-j)!} \, s_{n+j}}
\, , \qquad k, n \in \mathbb{N}_0 \, ,
% \label
\end{align}
and its recursive scheme \cite[Eq.\ (9.4-12)]{Weniger/1989}
\begin{subequations}
\label{linMinTrRecScheme}
\begin{align}
\mathcal{P}_0^{(n)} (\zeta, s_n) & \; = \; s_n \, ,
\qquad n \in \mathbb{N}_0 \, ,
% \label{}
\\
\mathcal{P}_{k+1}^{(n)} (\zeta, s_n) & \; = \;
\mathcal{P}_k^{(n+1)} (\zeta, s_{n+1}) \, + \,
\frac {\zeta+n-k} {k+1} \, \Delta \mathcal{P}_k^{(n)} (\zeta, s_n)
\, , \qquad k, n \in \mathbb{N}_0 \, .
% \label{}
\end{align}  
\end{subequations}

If we choose in (\ref{linCiZaSkaTr}) and (\ref{linCiZaSkaTrRecScheme})
$q_m = \chi + m/\alpha$, we obtain the Richardson-type variant of the
sequence transformation (\ref{GenCizTr}),
\begin{align}
{\vphantom{\mathcal{C}}}_{R} \mathcal{C}_k^{(n)} (\alpha, \chi, s_n) 
& \; = \; 
{\vphantom{\mathcal{G}}}_{R} \mathcal{G}_{k}^{(n)} (\chi+m/\alpha, s_n) 
\; = \; {\vphantom{\mathcal{C}}}_{R} 
\mathcal{C}_k^{(n)} \bigl( \alpha, \chi+1, s_n, 1/(\chi+n) \bigr)
\nonumber \\
& \; = \; 
{\frac{(-1)^k}{\alpha^k} \, \displaystyle \sum_{j=0}^k \,
(-1)^j \, \frac {(\alpha[\chi+n+j])_k} {j! (k-j)!} \, s_{n+j}}
\, , \qquad k, n \in \mathbb{N}_0 \, ,
\label{linCizTr}
\end{align}
and its recursive scheme
\begin{subequations}
\label{linCizTrRecScheme}
\begin{align}
{\vphantom{\mathcal{C}}}_{R} \mathcal{C}_0^{(n)} (\alpha, \chi, s_n) 
& \; = \; s_n \, , \qquad n \in \mathbb{N}_0 \, ,
% \label{}
\\
{\vphantom{\mathcal{C}}}_{R} \mathcal{C}_{k+1}^{(n)} (\alpha, \chi, s_n) 
& \; = \;
{\vphantom{\mathcal{C}}}_{R} \mathcal{C}_k^{(n+1)} (\alpha, \chi, s_{n+1}) 
\, + \, \frac {\chi+n+k/\alpha} {k+1} \, \Delta 
{\vphantom{\mathcal{C}}}_{R} \mathcal{C}_k^{(n)} (\alpha, \chi, s_n)
\, , \quad k, n \in \mathbb{N}_0 \, .
% \label{}
\end{align}  
\end{subequations}

Depending upon the value of $\alpha > 0$, ${\vphantom{\mathcal{C}}}_{R}
\mathcal{C}_k^{(n)} (\alpha, \chi, s_n)$ interpolates between the
Richardson-type transformations $\Lambda_k^{(n)} (\beta, s_n)$ and
$\mathcal{F}_k^{(n)} (\chi, s_n)$. If we choose in (\ref{linCizTr})
$\alpha = 1$ and compare the resulting expression with (\ref{linWenTr}),
we find
\begin{equation}
{\vphantom{\mathcal{C}}}_{R} \mathcal{C}_k^{(n)} (1, \chi, s_n) \; = \; 
\mathcal{F}_k^{(n)} (\chi, s_n) \, ,
% \label{}
\end{equation}
and if we use in (\ref{linCizTr}) 
\begin{equation}
\lim_{\alpha \to \infty} \, 
\frac{(\alpha[\chi + n + j])_{k}}{\alpha^{k}} 
\; = \; \lim_{\alpha \to \infty} \, 
\prod_{m=0}^{k-1} \, (\chi + n + j + m/\alpha)
\; = \; (\chi + n + j)^{k}
% \label{}
\end{equation}
together with $\chi = \beta$ and compare the resulting expression with
(\ref{RichAlLev}), we find
\begin{equation}
\lim_{\alpha \to \infty} \, 
{\vphantom{\mathcal{C}}}_{R} \mathcal{C}_k^{(n)} (\alpha, \beta, s_n)
\; = \; \Lambda_k^{(n)} (\beta, s_n) \, .
% \label{}
\end{equation}

\typeout{==> Rational Approximants}
\section{Rational Approximants}
\label{Sec:RatApprox}

In theoretical physics and in applied mathematics, Pad\'{e} approximants
\cite{Pade/1892} have become the standard tool to overcome problems with
slowly convergent or divergent power series. Pad\'{e} approximants can
also be viewed as a special class of sequence transformations since they
transform the partial sums
\begin{equation}
f_n (z) \; = \; \sum_{\nu=0}^{n} \, \gamma_{\nu} \, z^{\nu} \, , 
\qquad n \in \mathbb{N}_0 \, ,
\label{Par_Sum_PS}
\end{equation}
of a (formal) power series for some function $f$ into a doubly indexed
sequence of rational functions (see for example
\cite[Chapter 1]{Baker/Graves-Morris/1996}):
\begin{equation}
[ l / m ]_f (z) \; = \; \frac{P^{[l/m]} (z)}{Q^{[l/m]} (z)} \; = \;
\frac{p_0 + p_1 z + \ldots + p_{l} z^{l}}
{1 + q_1 z + \ldots + q_m z^m} \, , 
\qquad l, m \in \mathbb{N}_0 \, .
\label{DefPade}
\end{equation}

We also obtain rational approximants if the $u$, $t$, $d$, and $v$
variants considered in Section \ref{Sec:LevExpRemEst} are applied to the
partial sums (\ref{Par_Sum_PS}). Nevertheless, there are some
substantial differences between most sequence transformations and
Pad\'{e} approximants. Levin-type transformations can be computed via
their explicit expressions, although it is normally preferable to
compute them recursively. The coefficients $p_0$, \ldots, $p_l$ and
$q_1$, \ldots, $q_m$ of the two Pad\'{e} polynomials $P^{[l/m]}$ and
$Q^{[l/m]}$ in (\ref{DefPade}) are, however, chosen in such a way that
the Taylor expansion of the ratio $P^{[l/m]} (z) / Q^{[l/m]} (z)$ at $z
= 0$ agrees with the power series for $f$ as far as possible:
\begin{equation}
Q^{[l/m]} (z) \, f (z) \, - \, P^{[l/m]} (z) \; = \;
\mathrm{O} (z^{l + m + 1}) \, , \qquad z \to 0 \, .
\label{O_est_Pade}
\end{equation}
This asymptotic condition leads to a system of $l + m + 1$ linear
equations. If this system of equations has a solution, it yields the
coefficients of the polynomials $P^{[l/m]} (z)$ and $Q^{[l/m]} (z)$ (see
for example \cite[Chapter 1]{Baker/Graves-Morris/1996}).

In most practical applications, Pad\'{e} approximants are not computed
via the defining system of equations, but with the help of recursive
algorithms as for example Wynn's epsilon algorithm
\cite{Wynn/1956a}. Nevertheless, the accuracy-through-order relationship
(\ref{O_est_Pade}) guarantees that the Taylor expansion of $[ l / m ]_f
(z)$ reproduces the partial sum $f_{l+m} (z)$ from which it was
constructed. If a sequence transformation is applied to the partial sums
of a (formal) power series, it is by no means obvious whether the
resulting expression satisfies an accuracy-through-order relationship of
the type of (\ref{O_est_Pade}) (see for example the discussion in
\cite{Weniger/2000b}).

The accuracy-through-order relationship (\ref{O_est_Pade}) is essential
if Pad\'{e} approximants are to be used for the prediction of unknown
series coefficients, which was first described and utilized by Gilewicz
\cite{Gilewicz/1973}. This so-called Pad\'{e} prediction is based on the
fact that a Pad\'{e} approximant is by construction analytic at the
origin. Accordingly, the power series
\begin{equation}
[l/m]_f (z) \; = \;
\sum_{\nu=0}^{\infty} \, \gamma^{[l/m]}_{\nu} \, z^{\nu}
% \label{}
\end{equation}
converges in a neighborhood of $z=0$. The accuracy-through-order
relationship (\ref{O_est_Pade}) implies $\gamma^{[l/m]}_{\nu} =
\gamma_{\nu}$ for $0 \le \nu \le l+m$. The remaining coefficients
$\gamma^{[l/m]}_{l+m+\mu+1}$ with $\mu \ge 0$ are in general different
from the corresponding coefficients $\gamma_{l+m+\mu+1}$ of the power
series for $f (z)$. If, however, the Pad\'{e} approximants $[l/m]_f (z)$
converge more rapidly to $f (z)$ than the partial sums $f_{l+m} (z)$,
from which they are constructed, then the coefficients
$\gamma^{[l/m]}_{l+m+\mu+1}$ provide in particularly for smaller values
of $\mu$ approximants to the corresponding series coefficients. It is
important to note that Pad\'{e} prediction is not restricted to
convergent power series. Thus, Pad\'{e} prediction can produce
useful results even if the power series is a factorially divergent
perturbation expansion.

In certain subfields of theoretical physics, the computation of more
than a few coefficients of a perturbation expansion can be extremely
difficult. Moreover, these coefficients are often affected by
comparatively large relative errors. Under such adverse conditions,
Pad\'{e} approximants can be used to make predictions about the leading
unknown coefficients of perturbation expansions as well as to make
consistency checks for previously calculated coefficients. Further
details as well as many examples can be found in
\cite{Bender/Weniger/2001,Brodsky/Ellis/Gardi/Karliner/Samuel/1998,%
Chishtie/Elias/2001,Chishtie/Elias/Miransky/Steele/2000,%
Chishtie/Elias/Steele/1999a,Chishtie/Elias/Steele/1999b,%
Chishtie/Elias/Steele/2000a,Chishtie/Elias/Steele/2000b,%
Chishtie/Elias/Steele/2001,Cvetic/Yu/2000,%
Elias/Steele/Chishtie/Migneron/Sprague/1998,%
Ellis/Gardi/Karliner/Samuel/1996a,Ellis/Gardi/Karliner/Samuel/1996b,%
Ellis/Jack/Jones/Karliner/Samuel/1998,Ellis/Karliner/Samuel/1997,%
Jentschura/Becher/Weniger/Soff/2000,Jentschura/Weniger/Soff/2000,%
Karliner/1998,Samuel/Abraha/Yu/1997,Samuel/Ellis/Karliner/1995,%
Samuel/Li/1994,Samuel/Li/Steinfelds/1993,Samuel/Li/Steinfelds/1994,%
Samuel/Li/Steinfelds/1995,Samuel/Li/Steinfelds/1997,Steele/Elias/1998,%
Weniger/2000b} and in references therein. Pad\'{e} prediction can also
be quite helpful in different contexts. For example, Pad\'{e} prediction
techniques developed in \cite{Weniger/2000b} were used in
\cite{Bender/Weniger/2001} to provide numerical evidence that the
factorially divergent perturbation expansion for an anharmonic
oscillator, whose Hamiltonian is non-Hermitian but
$\mathcal{PT}$-symmetric \cite{Bender/Dunne/1999}, is a Stieltjes
series.

The prediction of unknown power series coefficients is not restricted to
Pad\'{e} approximants. In principle, any other rational approximant,
that also satisfies an accuracy-through-order relationship of the type
of (\ref{O_est_Pade}), can be used. It seems that this idea was first
formulated by Sidi and Levin \cite{Sidi/Levin/1983} and by Brezinski
\cite{Brezinski/1985b}. Recently, it was found that Levin-type
transformation like (\ref{dLevTr}) and (\ref{dWenTr}), which satisfy for
$k \in \mathbb{N}$ and $n \in \mathbb{N}_0$ the following asymptotic
order estimates as $z \to 0$
\cite[Eqs.\ (4.28) and (4.29)]{Weniger/Cizek/Vinette/1993},
\begin{align}
f (z) \, - \, d_{k}^{(n)} \bigl(\beta, f_n (z) \bigr) & \; = \;
\mathrm{O} \bigl( z^{k+n+2} \bigr) \, , 
\label{dLevTrPSOrdEst}
\\
f (z) \, - \, \delta_{k}^{(n)} \bigl(\beta, f_n (z) \bigr) & \; = \;
\mathrm{O} \bigl( z^{k+n+2} \bigr) \, ,
\label{dWenTrPSOrdEst}
\end{align}
produce at least in some cases significantly more accurate predictions
for unknown power series coefficients than Pad\'{e} approximants
\cite{Jentschura/Becher/Weniger/Soff/2000,Jentschura/Weniger/Soff/2000,%
Weniger/1997}. Accordingly. it should be of interest to analyze not only
the rational approximants, which result if Levin-type transformations
are applied to power series, but also their accuracy-through-order
relationships. In the case of Levin's $t$ transformation, this was
already done by Sidi and Levin \cite{Sidi/Levin/1983}, and the
accuracy-through-order relationships for the $u$, $t$, $v$, and $d$
variants of $\mathcal{L}_{k}^{(n)} (\beta, s_n, \omega_n)$, 
$\mathcal{S}_k^{(n)} (\beta, s_n, \omega_n)$, and 
$\mathcal{M}_k^{(n)} (\xi, s_n, \omega_n)$ were studied in 
\cite[Section 5.7]{Weniger/1994b}, albeit by a less elegant method.

In this Section, the $u$, $t$, $v$, and $d$ variants of the sequence
transformation $\mathcal{G}_{k}^{(n)} ( q_m, s_n, \omega_n)$ introduced
by \v{C}\'{\i}\v{z}ek, Zamastil, and Sk\'{a}la
\cite[Eq.\ (10)]{Cizek/Zamastil/Skala/2003} are applied to (formal)
power series and the accuracy-through-order properties of the resulting
rational approximants are studied. Since the sequence transformations
$\mathcal{L}_{k}^{(n)} (\beta, s_n, \omega_n)$, $\mathcal{S}_{k}^{(n)}
(\beta, s_n, \omega_n)$, $\mathcal{M}_{k}^{(n)} (\xi, s_n, \omega_n)$,
and $\mathcal{C}_{k}^{(n)} (\alpha, \beta, s_n, \omega_n)$ can be
obtained from $\mathcal{G}_{k}^{(n)} (q_m, s_n, \omega_n)$ by
specializing the parameters $q_m$, all results for
${\vphantom{\mathcal{G}}}_{u}
\mathcal{G}_{k}^{(n)} \bigl(q_m, f_n (z) \bigr)$, 
${\vphantom{\mathcal{G}}}_{t} 
\mathcal{G}_{k}^{(n)} \bigl(q_m, f_n (z) \bigr)$, 
${\vphantom{\mathcal{G}}}_{d} 
\mathcal{G}_{k}^{(n)} \bigl(q_m, f_n (z) \bigr)$, and 
${\vphantom{\mathcal{G}}}_{v} 
\mathcal{G}_{k}^{(n)} \bigl(q_m, f_n (z) \bigr)$ derived here can 
immediately be translated to the analogous results for the $u$, $t$,
$d$, and $v$ type variants of the transformations mentioned above.

If we use the partial sums (\ref{Par_Sum_PS}) of a (formal) power series
$f (z) = \sum_{\nu=0}^{n} \gamma_{\nu} z^{\nu}$ as input data, the
simple remainder estimates (\ref{GuRemEst}), (\ref{tRemEst}),
(\ref{vRemEst}), and (\ref{dRemEst}) for the $u$, $t$, $v$, and $d$
variants of $\mathcal{G}_{k}^{(n)} ( q_m, s_n, \omega_n)$ translate to
$\omega_n = (n+q_0) \gamma_n z^n$, $\omega_n = \gamma_n z^n$, $\omega_n
= \gamma_n \gamma_{n+1} z^{n+1}/[\gamma_n - z \gamma_{n+1}]$, and
$\omega_n = \gamma_{n+1} z^{n+1}$. Of course, these remainder estimates
can only be used if the coefficients of the power series for $f$ satisfy
$\gamma_n \ne 0$ for all $n \in \mathbb{N}_0$. In the following text,
this will be tacitly assumed.

If we apply the $u$ variant (\ref{uCiZaSkaTr}) of $\mathcal{G}_{k}^{(n)}
(q_m, s_n, \omega_n)$ to the partial sums (\ref{Par_Sum_PS}) of the
(formal) power series for $f$, we obtain:
\begin{eqnarray}
\lefteqn{{\vphantom{\mathcal{G}}}_{u} 
\mathcal{G}_{k}^{(n)} \bigl(q_m, f_n (z) \bigr) 
\; = \; 
\mathcal{G}_{k}^{(n)} \bigl( q_m, f_n (z), (n+q_0) \gamma_n z^n \bigr)} 
\nonumber \\
& = \; \frac
{\displaystyle
\sum_{j=0}^{k} \, (-1)^{j} \, {\binom{k}{j}} \,
\prod_{m=1}^{k-1} \, \frac {(n+j+q_m)}{(n+k+q_m)} \,
\frac {z^{k-j} f_{n+j} (z)} {(n+j+q_0) \gamma_{n+j}} }
{\displaystyle
\sum_{j=0}^{k} \, (-1)^{j} \, {\binom{k}{j}} \,
\prod_{m=1}^{k-1} \, \frac {(n+j+q_m)}{(n+k+q_m)} \,
\frac {z^{k-j}} {(n+j+q_0) \gamma_{n+j}} } 
\, , \qquad k, n \in \mathbb{N}_0 \, . 
\label{PSuCiZaSkaTr}
\end{eqnarray}

In the case of the $t$ variant (\ref{tCiZaSkaTr}) of
$\mathcal{G}_{k}^{(n)} (q_m, s_n, \omega_n)$, we obtain:
\begin{eqnarray}
\lefteqn{{\vphantom{\mathcal{G}}}_{t} 
\mathcal{G}_{k}^{(n)} \bigl(q_m, f_n (z) \bigr) 
\; = \; 
\mathcal{G}_{k}^{(n)} \bigl( q_m, f_n (z), \gamma_n z^n \bigr)} 
\nonumber \\
& = \; \frac
{\displaystyle
\sum_{j=0}^{k} \, (-1)^{j} \, {\binom{k}{j}} \,
\prod_{m=1}^{k-1} \, \frac {(n+j+q_m)}{(n+k+q_m)} \,
\frac {z^{k-j} f_{n+j} (z)} {\gamma_{n+j}} }
{\displaystyle
\sum_{j=0}^{k} \, (-1)^{j} \, {\binom{k}{j}} \,
\prod_{m=1}^{k-1} \, \frac {(n+j+q_m)}{(n+k+q_m)} \,
\frac {z^{k-j}} {\gamma_{n+j}} } 
\, , \qquad k, n \in \mathbb{N}_0 \, . 
\label{PStCiZaSkaTr}
\end{eqnarray}

In the case of the $d$ variant (\ref{dCiZaSkaTr}) of
$\mathcal{G}_{k}^{(n)} (q_m, s_n, \omega_n)$, we obtain:
\begin{eqnarray}
\lefteqn{{\vphantom{\mathcal{G}}}_{d} 
\mathcal{G}_{k}^{(n)} \bigl(q_m, f_n (z) \bigr) 
\; = \; 
\mathcal{G}_{k}^{(n)} \bigl( q_m, f_n (z), \gamma_{n+1} z^{n+1} \bigr)} 
\nonumber \\
& = \; \frac
{\displaystyle
\sum_{j=0}^{k} \, (-1)^{j} \, {\binom{k}{j}} \,
\prod_{m=1}^{k-1} \, \frac {(n+j+q_m)}{(n+k+q_m)} \,
\frac {z^{k-j} f_{n+j} (z)} {\gamma_{n+j+1}} }
{\displaystyle
\sum_{j=0}^{k} \, (-1)^{j} \, {\binom{k}{j}} \,
\prod_{m=1}^{k-1} \, \frac {(n+j+q_m)}{(n+k+q_m)} \,
\frac {z^{k-j}} {\gamma_{n+j+1}} } 
\, , \qquad k, n \in \mathbb{N}_0 \, . 
\label{PSdCiZaSkaTr}
\end{eqnarray}

The numerators of (\ref{PSuCiZaSkaTr}) - (\ref{PSdCiZaSkaTr}) are
polynomials of degree $k+n$ in $z$, and the denominators are polynomials
of degree $k$ in $z$. For the computation of (\ref{PSuCiZaSkaTr}) and
(\ref{PStCiZaSkaTr}), we need the numerical values of the partial sums
$f_{n} (z)$, $f_{n+1} (z)$, $\ldots$, $f_{n+k} (z)$, and for the
computation of (\ref{PSdCiZaSkaTr}), we need the partial sums $f_{n}
(z)$, $f_{n+1} (z)$, $\ldots$, $f_{n+k+1} (z)$.

In the case of the $v$ variant (\ref{vCiZaSkaTr}) of
$\mathcal{G}_{k}^{(n)} (q_m, s_n, \omega_n)$, we obtain:
\begin{eqnarray}
\lefteqn{{\vphantom{\mathcal{G}}}_{v} 
\mathcal{G}_{k}^{(n)} \bigl(q_m, f_n (z) \bigr) 
\; = \; 
\mathcal{G}_{k}^{(n)} \bigl( q_m, f_n (z), \gamma_n \gamma_{n+1} z^{n+1}
/[\gamma_n - z \gamma_{n+1}]\bigr)} 
\nonumber \\
& = \; \frac
{\displaystyle
\sum_{j=0}^{k} \, (-1)^{j} \, {\binom{k}{j}} \,
\prod_{m=1}^{k-1} \, \frac {(n+j+q_m)}{(n+k+q_m)} \,
\frac {z^{k-j} (\gamma_{n+j} - z \gamma_{n+j+1}) f_{n+j} (z)} 
{\gamma_{n+j} \gamma_{n+j+1}} }
{\displaystyle
\sum_{j=0}^{k} \, (-1)^{j} \, {\binom{k}{j}} \,
\prod_{m=1}^{k-1} \, \frac {(n+j+q_m)}{(n+k+q_m)} \,
\frac {z^{k-j} (\gamma_{n+j} - z \gamma_{n+j+1})} 
{\gamma_{n+j} \gamma_{n+j+1}} } 
\, , \quad k, n \in \mathbb{N}_0 \, . \quad
\label{PSvCiZaSkaTr}
\end{eqnarray}
The numerator of this expression is a polynomial of degree $k+n+1$ in
$z$, and the denominator is a polynomial of degree $k+1$ in $z$. For its
computation, we need the numerical values of the partial sums $f_{n}
(z)$, $f_{n+} (z)$, $\ldots$, $f_{n+k+1} (z)$.

Next, asymptotic order estimates of the type of (\ref{O_est_Pade}) will
be constructed for the rational approximants (\ref{PSuCiZaSkaTr}) -
(\ref{PSvCiZaSkaTr}). Here, it must be taken into account that an
accuracy-through-order relationship does not make any sense if the
rational function reproduces exactly the function $f$ represented by the
power series. This is for instance the case if $u$, $t$, $d$, and $v$
variants of $\mathcal{G}_{k}^{(n)} (q_m, s_n, \omega_n)$ are applied to
the partial sums $\sum_{\nu=0}^{n} z^{\nu} = (1-z^{n+1})/(1-z)$ of the
geometric series. For an analysis of these complications, let us
consider the $u$ variant (\ref{PSuCiZaSkaTr}). If we introduce the
remainders of the partial sums (\ref{Par_Sum_PS}) according to
\begin{equation}
r_n (z) \; = \; f_n (z) \, - \, f (z) \; = \; -
\sum_{\nu = 0}^{\infty} \, \gamma_{n + \nu + 1} \, z^{n + \nu + 1} \, ,
\label{DefRenPS}
\end{equation}
then the difference between $f$ and the $u$ variant (\ref{PSuCiZaSkaTr})
can according to (\ref{GenLevTypeTrDiffOpRepLimPlusRn}) be expressed as
follows:
\begin{eqnarray}
\lefteqn{f (z) \, - \, {\vphantom{\mathcal{G}}}_{u}
\mathcal{G}_{k}^{(n)} \bigl(q_m, f_n (z) \bigr)}
\nonumber \\
& = \; - \, z^{k+n} \, \frac
{\displaystyle
\sum_{j=0}^{k} \, (-1)^{j} \, {\binom{k}{j}} \,
\prod_{m=1}^{k-1} \, (n+j+q_m) \,
\frac {r_{n+j} (z)} {(n+j+q_0) \gamma_{n+j} z^{n+j}} }
{\displaystyle
\sum_{j=0}^{k} \, (-1)^{j} \, {\binom{k}{j}} \,
\prod_{m=1}^{k-1} \, (n+j+q_m) \,
\frac {z^{k-j}} {(n+j+q_0) \gamma_{n+j}} } \, .
\label{PS_Rem_uCiZaSkaTr}
\end{eqnarray}
The denominator of this expression is by assumption of order $\mathrm{O}
(1)$ as $z \to 0$. For the derivation of an order estimate of the
numerator, we use (\ref{DeltaPk}) to obtain
\begin{eqnarray}
\lefteqn{\sum_{j=0}^{k} \, (-1)^{j} \, {\binom{k}{j}} \,
\prod_{m=1}^{k-1} \, (n+j+q_m) \,
\frac {r_{n+j} (z)} {(n+j+q_0) \gamma_{n+j} z^{n+j}}} \nonumber \\
& = \; (-1)^k \, {\displaystyle
\Biggl[ \Delta^k \, \prod_{m=1}^{k-1} \, (n+q_m) \,
\frac {r_{n} (z)} {(n+q_0) \gamma_{n} z^{n}} \Biggr]} \, .
\label{NumerEst1}
\end{eqnarray}
We now have to distinguish some special cases. 

Let us first assume that the $u$ type remainder estimate $\omega_n =
(n+q_0) \gamma_n z^n$ is a perfect remainder estimate according to
(\ref{SeqPerfecRemEst}). Then, $\Delta^k$ acts on a polynomial of degree
$k-1$ in $n$, which means that the right-hand side of
(\ref{PS_Rem_uCiZaSkaTr}) is annihilated. Accordingly,
${\vphantom{\mathcal{G}}}_{u} \mathcal{G}_{k}^{(n)} \bigl(q_m, f_n (z)
\bigr)$ is exact for $k \ge 1$ and an accuracy-through-order
relationship of the type of (\ref{O_est_Pade}) makes no sense.

Let us now assume that the $t$ type remainder estimate $\omega_n =
\gamma_n z^n$ is a perfect remainder estimate according to
(\ref{SeqPerfecRemEst}), which is the case if the input data $\{ f_n (z)
\}_{n=0}^{\infty}$ are the partial sums $\sum_{\nu=0}^{n} z^{\nu} =
(1-z^{n+1})/(1-z)$ of the geometric series. Then, we again have to
distinguish two cases. If $q_0 \in \{ q_1, \ldots, q_m \}$, the ratio
$\prod_{m=1}^{k-1} (n+q_m)/(n+q_0)$ simplifies to yield a polynomial
of degree $k-2$ in $n$ which is annihilated by $\Delta^k$. Accordingly,
${\vphantom{\mathcal{G}}}_{u} \mathcal{G}_{k}^{(n)} \bigl(q_m, f_n (z)
\bigr)$ is for $k \ge 1$ exact for the partial sums of the geometric
series and an accuracy-through-order relationship makes no sense. If
$q_0 \notin \{ q_1, \ldots, q_m \}$, the ratio $\prod_{m=1}^{k-1}
(n+q_m)/ (n+q_0)$ does not simplify to yield a polynomial and is not
annihilated by $\Delta^k$. Accordingly, ${\vphantom{\mathcal{G}}}_{u}
\mathcal{G}_{k}^{(n)} \bigl(q_m, f_n (z) \bigr)$ is in this case not
exact for the geometric series.

The exactness for the geometric series is probably the most fundamental
requirement for a sequence transformation in the case of linear
convergence ($0 < \vert \rho \vert < 1$ in (\ref{DefLinLogConv})). This
follows from Germain-Bonne's formal theory of convergence acceleration
\cite{Germain-Bonne/1973} and its extension to Levin-type
transformations \cite[Section 12]{Weniger/1989}. Consequently, it is
probably a good idea that to choose $q_0$ in in $\omega_n = (n+q_0)
\Delta s_{n-1}$ according to $q_0 \in \{ q_1, \ldots, q_m \}$. An 
obvious idea would be to choose $q_0 = q_1$. The remainder estimates of
the other $u$ type transformations (\ref{uLevTr}), (\ref{uWenTr}),
(\ref{uMinTr}), and (\ref{uCizTr}) all satisfy this
requirement. Accordingly, these $u$ variants are for $k \ge 1$ exact for
the geometric series. This is also true for the $t$, $d$, and $v$
variants (\ref{PStCiZaSkaTr}) - (\ref{PSvCiZaSkaTr}) of
$\mathcal{G}_{k}^{(n)} \bigl(q_m, s_n, \omega_n \bigr)$.

By analyzing expressions of the type of (\ref{PS_Rem_uCiZaSkaTr}),
accuracy-through-order relationships for the rational approximants
${\vphantom{\mathcal{G}}}_{u} \mathcal{G}_{k}^{(n)}
\bigl(q_m, f_n (z) \bigr)$, ${\vphantom{\mathcal{G}}}_{u} 
\mathcal{G}_{k}^{(n)} \bigl(q_m, f_n (z) \bigr)$, 
${\vphantom{\mathcal{G}}}_{u} \mathcal{G}_{k}^{(n)} 
\bigl(q_m, f_n (z) \bigr)$, and ${\vphantom{\mathcal{G}}}_{u} 
\mathcal{G}_{k}^{(n)} \bigl(q_m, f_n (z) \bigr)$ can be
derived (see for example \cite[Section 5.7]{Weniger/1994b}). However, it
is more elegant to do this via the theory of Pad\'{e}-type approximants
\cite{Brezinski/1980a}. 

As is well known, the coefficients of the numerator and the denominator
polynomials of a Pad\'{e} approximants are chosen in such a way that the
asymptotic order estimate (\ref{O_est_Pade}) is satisfied, but it is not
so well known that generalizations and modifications of Pad\'{e}
approximants can be obtained by suitably modifying the asymptotic
condition (\ref{O_est_Pade}). For example, let us consider the rational
approximants
\begin{equation}
( l / m )_f (z) \; = \; 
\frac{\mathcal{U}^{(l/m)} (z)}{\mathcal{V}^{(l/m)} (z)} \; = \;
\frac{u_0 + u_1 z + \ldots + u_{l} z^{l}}
{v_0 + v_1 z + \ldots + v_m z^m} \, , 
\qquad l, m \in \mathbb{N}_0 \, .
\label{DefPadeType}
\end{equation}
We assume that the two polynomials $\mathcal{U}^{(l/m)} (z)$ and
$\mathcal{V}^{(l/m)} (z)$ are exactly of degrees $l$ and $m$ in $z$, or
equivalently that $u_l \ne 0$ and $v_m \ne 0$. Let us now assume that
the $m+1$ coefficients $v_0$, $v_1$, \ldots, $v_m$ of the denominator
polynomial $\mathcal{V}^{(l/m)} (z)$ are chosen according to some
rule. Then, only the $l+1$ coefficients $u_0$, $u_1$, \ldots, $u_l$ of
the numerator polynomial $\mathcal{U}^{(l/m)} (z)$ have to be
determined via the modified asymptotic condition
\begin{equation}
\mathcal{V}^{(l/m)} (z) \, f (z) \, - \, \mathcal{U}^{(l/m)} (z) \; = \;
\mathrm{O} (z^{l + 1}) \, , \qquad z \to 0 \, ,
\label{O_est_PadeType}
\end{equation}
yielding
\begin{equation}
\mathcal{U}^{(l/m)} (z) \; = \; 
\sum_{\lambda=0}^{l} \, v_{\lambda} \, z^{\lambda} \, f_{l-\lambda} (z)
\, .
\label{DefNumPadeType}
\end{equation}

The rational function $( l / m )_f (z)$ is a so-called Pad\'{e}-type
approximant. Pad\'{e}-type approximants and their properties are
discussed in a monograph by Brezinski \cite{Brezinski/1980a}.

Let us now set $m = k$ and $l = k+n$ with $k, n, \in \mathbb{N}_0$ in
(\ref{DefPadeType}). Then, (\ref{DefNumPadeType}) implies
\begin{equation}
(k+n/k)_f (z) \; = \;
\frac{\mathcal{U}^{(k+n/k)} (z)}{\mathcal{V}^{(k+n/k)} (z)} \; = \;
\frac{\displaystyle 
\sum_{j=0}^{k} \, v_j \, z^j \, f_{k+n-j} (z)}
{\displaystyle 
\sum_{j=0}^{k} \, v_j \, z^j} \; = \;
\frac{\displaystyle 
\sum_{j=0}^{k} \, v_{k-j} \, z^{k-j} \, f_{n+j} (z)}
{\displaystyle 
\sum_{j=0}^{k} \, v_{k-j} \, z^{k-j}} \, .
% \label{}
\end{equation}
It follows from (\ref{PSuCiZaSkaTr}), (\ref{PStCiZaSkaTr}), and
(\ref{PSdCiZaSkaTr}) that the rational approximants
${\vphantom{\mathcal{G}}}_{u}
\mathcal{G}_{k}^{(n)} \bigl(q_m, f_n (z) \bigr)$, 
${\vphantom{\mathcal{G}}}_{t} 
\mathcal{G}_{k}^{(n)} \bigl(q_m, f_n (z) \bigr)$, 
and ${\vphantom{\mathcal{G}}}_{d}
\mathcal{G}_{k}^{(n)} \bigl(q_m, f_n (z) \bigr)$ possess the following 
general structure:
\begin{equation}
\mathbf{T}_{k}^{(n)} (z) \; = \; \frac
{\displaystyle 
\sum_{j=0}^{k} \, \lambda_{j}^{(k, n)} \, z^{k-j} \, f_{n+j} (z)} 
{\displaystyle 
\sum_{j=0}^{k} \, \lambda_{j}^{(k, n)} \, z^{k-j}} 
 \; = \; \frac
{\displaystyle 
\sum_{j=0}^{k} \, \lambda_{k-j}^{(k, n)} \, z^j \, f_{n+k-j} (z)} 
{\displaystyle 
\sum_{j=0}^{k} \, \lambda_{k-j}^{(k, n)} \, z^j} \, , 
\qquad k, n \in \mathbb{N}_0 \, .
\label{DefTkn}
\end{equation}
Thus, ${\vphantom{\mathcal{G}}}_{u}
\mathcal{G}_{k}^{(n)} \bigl(q_m, f_n (z) \bigr)$, 
${\vphantom{\mathcal{G}}}_{t} 
\mathcal{G}_{k}^{(n)} \bigl(q_m, f_n (z) \bigr)$, 
and ${\vphantom{\mathcal{G}}}_{d}
\mathcal{G}_{k}^{(n)} \bigl(q_m, f_n (z) \bigr)$ are Pad\'{e}-type 
approximants of the type of $(k+n/k)_f (z)$ with $v_{j} =
\lambda_{k-j}^{(k, n)}$.

It is a direct consequence of the defining asymptotic condition
(\ref{O_est_PadeType}) that the Pad\'{e}-type approximant
$\mathbf{T}_{k}^{(n)} (z)$ satisfies for all $k, n \in \mathbb{N}_0$ the
accuracy-through-order relationship
\begin{equation}
f (z) \, - \, \mathbf{T}_{k}^{(n)} (z) \; = \; 
\mathrm{O} \bigl( z^{k+n+1} \bigr) \, , \qquad z \to 0 \, .
\label{AsyEst_Tkn}
\end{equation}
This implies that the functions ${\vphantom{\mathcal{G}}}_{u} 
\mathcal{G}_{k}^{(n)} \bigl(q_m, f_n (z) \bigr)$ and 
${\vphantom{\mathcal{G}}}_{t} 
\mathcal{G}_{k}^{(n)} \bigl(q_m, f_n (z) \bigr)$ as well as all other
$u$ and $t$ type transformations considered in this article satisfy for
$k, n \in \mathbb{N}_0$ the following asymptotic order estimates as $z
\to \infty$:
\begin{align}
f (z) \, - \, {\vphantom{\mathcal{G}}}_{u} 
\mathcal{G}_{k}^{(n)} \bigl(q_m, f_n (z) \bigr) \; = \; 
\mathrm{O} \bigl( z^{k+n+1} \bigr) \, ,
\label{AsyEst_u_type} \\
f (z) \, - \, {\vphantom{\mathcal{G}}}_{t} 
\mathcal{G}_{k}^{(n)} \bigl(q_m, f_n (z) \bigr) \; = \; 
\mathrm{O} \bigl( z^{k+n+1} \bigr) \, .
\label{AsyEst_t_type}
\end{align}
Thus, all coefficients $\gamma_0$, $\gamma_1$, \ldots, $\gamma_{k+n}$ of
the power series $f (z) = \sum_{\nu=0}^{\infty} \gamma_{\nu} z^{\nu}$,
that are used for the construction of the rational approximants
${\vphantom{\mathcal{G}}}_{u} \mathcal{G}_{k}^{(n)} \bigl(q_m, f_n (z)
\bigr)$ and ${\vphantom{\mathcal{G}}}_{t} \mathcal{G}_{k}^{(n)}
\bigl(q_m, f_n (z) \bigr)$, respectively, are reproduced by a Taylor 
expansion around $z = 0$.

For ${\vphantom{\mathcal{G}}}_{d} \mathcal{G}_{k}^{(n)}
\bigl(q_m, f_n (z) \bigr)$ we obtain the same asymptotic order estimate:
\begin{equation}
f (z) \, - \, {\vphantom{\mathcal{G}}}_{d}
\mathcal{G}_{k}^{(n)} \bigl(q_m, f_n (z) \bigr) \; = \;
\mathrm{O} \bigl( z^{k+n+1} \bigr) \, , 
\qquad k, n \in \mathbb{N}_0 \, , \quad z \to 0 \, ,
\label{WrongAsyEst_d_type}
\end{equation}

In the context of the prediction of unknown power series coefficients,
this is a highly unwelcome result: For the computation of
${\vphantom{\mathcal{G}}}_{d} \mathcal{G}_{k}^{(n)}
\bigl(q_m, f_n (z) \bigr)$ we need the series coefficients $\gamma_0$, 
$\gamma_1$, \ldots, $\gamma_{n+k+1}$. Thus, the order term $\mathrm{O}
\bigl( z^{k+n+1} \bigr)$ implies that a Taylor expansion of
${\vphantom{\mathcal{G}}}_{d} \mathcal{G}_{k}^{(n)}
\bigl(q_m, f_n (z) \bigr)$ does not reproduce all coefficients
used for its construction. Moreover, the order estimates
(\ref{dLevTrPSOrdEst}) and (\ref{dWenTrPSOrdEst}), which were derived
in \cite{Weniger/Cizek/Vinette/1993} by directly analyzing the
corresponding expressions without using the theory of Pad\'{e}-type
approximants, indicate that we should instead get the order estimate
\begin{equation}
f (z) \, - \, {\vphantom{\mathcal{G}}}_{d}
\mathcal{G}_{k}^{(n)} \bigl(q_m, f_n (z) \bigr) \; = \;
\mathrm{O} \bigl( z^{k+n+2} \bigr) \, , \qquad k \in \mathbb{N} \, , 
\quad n \in \mathbb{N}_0 \, , \qquad z \to 0 \, .
\label{AsyEst_d_type}
\end{equation}

It is indeed possible to derive this seemingly irregular
accuracy-through-order relationship by analyzing the Pad\'{e}-type
approximant $\mathbf{T}_{k}^{(n)} (z)$ more carefully. For that purpose,
we rewrite (\ref{DefTkn}) as follows:
\begin{equation}
\mathbf{T}_{k}^{(n)} (z) \; = \; f (z) \, - \, z^{k+n+1} \, \frac
{\displaystyle 
\sum_{j=0}^{k} \, \lambda_{j}^{(k, n)} \,
\sum_{\nu=0}^{\infty} \, \gamma_{n+j+\nu+1} \, z^{\nu}} 
{\displaystyle 
\sum_{j=0}^{k} \, \lambda_{j}^{(k, n)} \, z^{k-j}} \, .
% \label{TknTrans_d}
\end{equation}
The denominator on the right-hand side is by assumption of order
$\mathrm{O} (1)$ as $z \to 0$. Accordingly, the asymptotic estimate
(\ref{AsyEst_Tkn}) is normally optimal, and the improved asymptotic
estimate (\ref{AsyEst_d_type}) can only hold if the $z$-independent part
of the numerator vanishes, or equivalently if $\sum_{j=0}^{k}
\lambda_{j}^{(k, n)} \gamma_{n+j+1} = 0$. For essentially arbitrary 
coefficients $\lambda_{j}^{(k, n)}$ this is certainly not true. However,
in the case of all $d$ type transformations of this article we have
$\lambda_{j}^{(k, n)} = (-1)^j {\binom{k}{j}} P_{k-1} (n+j) /
\gamma_{n+j+1}$, where $P_{k-1} (n)$ is a suitable polynomial of degree
$k-1$ in $n$. Then, we have for $k \ge 1$
\begin{equation}
\sum_{j=0}^{k} \, \lambda_{j}^{(k, n)} \, \gamma_{n+j+1} \; = \;
\sum_{j=0}^{k} \, (-1)^j \, {\binom{k}{j}} \, P_{k-1} (n+j) \; = \;
(-1)^k \, \Delta^k \, P_{k-1} (n) \; = \; 0 \, .
\label{VanishLambdaSum}
\end{equation}
This proofs the refined accuracy-through-order relationship
(\ref{AsyEst_d_type}).

It follows from (\ref{PSvCiZaSkaTr}) that the rational approximant
${\vphantom{\mathcal{G}}}_{v} \mathcal{G}_{k}^{(n)} 
\bigl(q_m, f_n (z) \bigr)$ possesses like all other $v$ type
transformations of this article the following general structure:
\begin{equation}
\mathbf{V}_{k}^{(n)} (z) \; = \; \frac
{\displaystyle 
\sum_{j=0}^{k} \, \lambda_{j}^{(k, n)} \, z^{k-j} \, f_{n+j} (z) \, + \,
z \, \sum_{j=0}^{k} \, \mu_{j}^{(k, n)} \, z^{k-j} \, f_{n+j} (z)} 
{\displaystyle 
\sum_{j=0}^{k} \, \lambda_{j}^{(k, n)} \, z^{k-j} \, + \,
z \, \sum_{j=0}^{k} \, \mu_{j}^{(k, n)} \, z^{k-j}} \, ,
\qquad k, n \in \mathbb{N}_0 \, .
\label{DefVkn}
\end{equation}
Comparison with (\ref{DefTkn}) shows that $\mathbf{V}_{k}^{(n)} (z)$ is
no Pad\'{e}-type approximant. Accordingly, the defining asymptotic
condition (\ref{O_est_PadeType}) of Pad\'{e}-type approximants cannot be
applied. Fortunately, an analogous asymptotic order estimate for
$\mathbf{V}_{k}^{(n)} (z)$ can be derived easily. For that purpose, we
rewrite (\ref{DefVkn}) as follows:
\begin{eqnarray}
\lefteqn{\mathbf{V}_{k}^{(n)} (z) \; = \; f (z)} \nonumber \\
& - \, z^{k+n+1} \, \frac
{\displaystyle 
\sum_{j=0}^{k} \, \lambda_{j}^{(k, n)} \, 
\sum_{\nu=0}^{\infty} \, \gamma_{n+j+\nu+1} \, z^{\nu} \, + \,
z \, \sum_{j=0}^{k} \, \mu_{j}^{(k, n)} \, 
\sum_{\nu=0}^{\infty} \, \gamma_{n+j+\nu+1} \, z^{\nu}} 
{\displaystyle 
\sum_{j=0}^{k} \, \lambda_{j}^{(k, n)} \, z^{k-j} \, + \,
z \, \sum_{j=0}^{k} \, \mu_{j}^{(k, n)} \, z^{k-j}} \, .
\label{VknTrans}
\end{eqnarray}
The denominator on the right-hand side is by assumption of order
$\mathrm{O} (1)$ as $z \to 0$. Accordingly, we obtain the
asymptotic order estimate
\begin{equation}
f (z) \, - \, \mathbf{V}_{k}^{(n)} (z) \; = \;
\mathrm{O} \bigl( z^{k+n+1} \bigr) \, ,
\qquad k, n \in \mathbb{N}_0 \, , \quad z \to 0 \, ,
\label{AsyEst_Vkn}
\end{equation}
which implies
\begin{equation}
f (z) \, - \, {\vphantom{\mathcal{G}}}_{v}
\mathcal{G}_{k}^{(n)} \bigl(q_m, f_n (z) \bigr) \; = \;
\mathrm{O} \bigl( z^{k+n+1} \bigr) \, ,
\qquad k, n \in \mathbb{N}_0 \, , \quad z \to 0 \, .
\label{WrongAsyEst_v_type}
\end{equation}

Now, we have the same problem as in the case of the suboptimal order
estimate (\ref{WrongAsyEst_d_type}) for ${\vphantom{\mathcal{G}}}_{d}
\mathcal{G}_{k}^{(n)} \bigl(q_m, f_n (z) \bigr)$: The order term 
$\mathrm{O} \bigl( z^{k+n+1} \bigr)$ in (\ref{WrongAsyEst_v_type})
implies that a Taylor expansion of ${\vphantom{\mathcal{G}}}_{v}
\mathcal{G}_{k}^{(n)} \bigl(q_m, f_n (z) \bigr)$ reproduces only
$\gamma_0$, $\gamma_1$, \ldots, $\gamma_{n+k}$, whereas $\gamma_0$, 
$\gamma_1$, \ldots, $\gamma_{n+k+1}$ are needed
for the computation of ${\vphantom{\mathcal{G}}}_{v}
\mathcal{G}_{k}^{(n)} \bigl(q_m, f_n (z) \bigr)$. Thus, we need instead 
the order estimate
\begin{equation}
f (z) \, - \, {\vphantom{\mathcal{G}}}_{v}
\mathcal{G}_{k}^{(n)} \bigl(q_m, f_n (z) \bigr) \; = \;
\mathrm{O} \bigl( z^{k+n+2} \bigr) \, , \qquad k \in \mathbb{N} \, , 
\quad n \in \mathbb{N}_0 \, , \quad z \to 0 \, .
\label{AsyEst_v_type}
\end{equation}
This refined asymptotic estimate can only be true if the
$z$-independent part of the first numerator sum in (\ref{VknTrans})
vanishes, or equivalently if $\sum_{j=0}^{k} \lambda_{j}^{(k, n)}
\gamma_{n+j+1} = 0$. For essentially arbitrary coefficients 
$\lambda_{j}^{(k, n)}$ this is certainly not true. However, in the case
of the $v$ type transformations of this article we have just like in the
case of the $d$ type transformations $\lambda_{j}^{(k, n)} = (-1)^j
{\binom{k}{j}} P_{k-1} (n+j) / \gamma_{n+j+1}$ and $\mu_{j}^{(k,n)} = 
(-1)^{j+1} {\binom{k}{j}} P_{k-1} (n+j) / \gamma_{n+j}$, where
$P_{k-1} (n)$ is a suitable polynomial of degree $k-1$ in $n$. Thus,
(\ref{VanishLambdaSum}) holds which proves the accuracy-through-order
relationship (\ref{AsyEst_v_type}).

In Section \ref{Sec:LevExpRemEst}, it was mentioned that in some cases
asymptotic expressions $a_{n}^{(\infty)}$ for the terms $a_n$ of an
infinite series $s = \sum_{\nu=0}^{\infty} a_{\nu}$ are known which
reproduce the leading order asymptotics of $a_n$ as $n \to \infty$, and
that these asymptotic expressions can also be used in the simple
remainder estimates (\ref{uRemEst}), (\ref{tRemEst}), (\ref{vRemEst}),
and (\ref{dRemEst}) since they also reproduce the leading order
asymptotics of the remainders $r_n = s_n-s$ as $n \to \infty$.

Thus, we now assume that asymptotic expressions $\gamma_{n}^{(\infty)}$
are known that reproduce the leading order asymptotics of the
coefficients of the power series $f (z) = \sum_{\nu=0}^{\infty} 
\gamma_{\nu} z^{\nu}$ according to
\begin{equation}
\gamma_n \; = \; \gamma_{n}^{(\infty)} \, 
\bigl[c + \mathrm{O} (1/n)  \bigr] \, ,
\qquad c \ne 0 \, , \quad n \to \infty \, .
% \label{}
\end{equation}
If we use in $\mathcal{G}_{k}^{(n)} ( q_m, s_n, \omega_n)$ the $u$, $t$,
$v$, and $d$ type remainder estimates $\omega_n = (n+q_0)
\gamma_n^{(\infty)} z^n$, $\omega_n = \gamma_n^{(\infty)} z^n$, 
$\omega_n = \gamma_n^{(\infty)} \gamma_{n+1}^{(\infty)} z^{n+1}/
[\gamma_n^{(\infty)} - z \gamma_{n+1}^{(\infty)}]$, and $\omega_n =  
\gamma_{n+1}^{(\infty)} z^{n+1}$, we obtain rational approximants which 
closely resemble the $u$, $t$, $v$, and $d$ variants
(\ref{PSuCiZaSkaTr}), (\ref{PStCiZaSkaTr}),(\ref{PSdCiZaSkaTr}), and
(\ref{PSvCiZaSkaTr}), and which are also special cases of the rational
functions $\mathbf{T}_{k}^{(n)} (z)$ and $\mathbf{V}_{k}^{(n)} (z)$
defined in (\ref{DefTkn}) and (\ref{DefVkn}),
respectively. Consequently, these rational approximants satisfy for all
$k, n \in \mathbb{N}_0$ the following asymptotic estimates as $z \to 0$:
\begin{align}
f (z) \, - \, \mathcal{G}_{k}^{(n)} \bigl( q_m, f_n (z), 
(n+q_0) \gamma_n^{(\infty)} z^n \bigr) & \; = \; 
\mathrm{O} \bigl( z^{k+n+1} \bigr) \, ,
% \label{}
\\
f (z) \, - \, \mathcal{G}_{k}^{(n)} \bigl( q_m, f_n (z), 
\gamma_n^{(\infty)} z^n \bigr) & \; = \;
\mathrm{O} \bigl( z^{k+n+1} \bigr) \, , 
% \label{}
\\
f (z) \, - \, \mathcal{G}_{k}^{(n)} \bigl( q_m, f_n (z), 
\gamma_{n+1}^{(\infty)} z^n \bigr) & \; = \; 
\mathrm{O} \bigl( z^{k+n+1} \bigr) \, ,
% \label{}
\\
f (z) \, - \, \mathcal{G}_{k}^{(n)} \bigl( q_m, f_n (z), 
\gamma_n^{(\infty)} \gamma_{n+1}^{(\infty)} z^{n+1}
/[\gamma_n^{(\infty)} - z \gamma_{n+1}^{(\infty)}]\bigr) & \; = \;
\mathrm{O} \bigl( z^{k+n+1} \bigr) \, .
\end{align}
Accordingly, all coefficients $\gamma_0$, $\gamma_1$, \ldots,
$\gamma_{k+n}$ of the power series for $f (z)$, which were used for the
construction of these rational approximants. are reproduced by Taylor
expansion.

Improved asymptotic estimates of the type of (\ref{AsyEst_d_type}) or
(\ref{AsyEst_v_type}) for the $d$ and $v$ type variants do not hold
here. The reason is that the coefficients $\lambda_{j}^{(k, n)}$ in
(\ref{DefTkn}) and (\ref{DefVkn}) now satisfy $\lambda_{j}^{(k, n)} =
(-1)^j {\binom{k}{j}} P_{k-1} (n+j) / \gamma_{n+j+1}^{(\infty)}$. In
general, we have $\gamma_{n+j+1}^{(\infty)} \ne \gamma_{n+j+1}$, which
implies that $\sum_{j=0}^{k} \lambda_{j}^{(k, n)} \gamma_{n+j+1} = 0$
does not hold.

If we use in $\mathcal{G}_{k}^{(n)} ( q_m, s_n, \omega_n)$ the remainder
estimates $\omega_n = (n+q_0) \gamma_n^{(\infty)} z^n$, $\omega_n = 
\gamma_n^{(\infty)} z^n$, $\omega_n = \gamma_n^{(\infty)} 
\gamma_{n+1}^{(\infty)} z^{n+1}/[\gamma_n^{(\infty)} - z 
\gamma_{n+1}^{(\infty)}]$, and $\omega_n =  \gamma_{n+1}^{(\infty)}
z^{n+1}$, the poles of the resulting rational approximants are
determined by the parameters $q_m$ in the products $\prod_{m=1}^{k-1}
(n+q_m)$ and by the remainder estimates, but they do not depend on the
coefficients $\gamma_n$ of the power series for $f$. This highlights
once more the crucial importance of the remainder estimates for the
success or the failure of Levin-type transformations. In contrast, both
the numerator and the denominator coefficients of a Pad\'{e} approximant
$[ l / m ]_f (z)$ depend via (\ref{O_est_Pade}) on the coefficients of
the partial sum $f_{l+m} (z)$ which was used for its construction.

\typeout{==> Summary and Outlook}
\section{Summary and Outlook}
\label{Sec:SummOutlook}

Levin \cite{Levin/1973} deserves credit for realizing that the
efficiency of convergence acceleration and summation processes can be
enhanced considerably by using as input data not only the elements of
the sequence $\{ s_n \}_{n=0}^{\infty}$ to be transformed, but also
explicit estimates $\{ \omega_n \}_{n=0}^{\infty}$ for the truncation
errors of this sequence.

If the input data are the partial sums of an infinite series, and if
sufficiently simple analytical expressions for the terms of this series
are known, then it is possible to derive analytical estimates for the
truncation errors of these series. In principle, the use of specially
designed analytical remainder estimates would be highly desirable,
although the resulting expressions are no longer generally applicable
sequence transformations, but rather (optimized) approximation schemes
for specific problems (see for instance the discussion in
\cite{Weniger/2001*}).

However, convergence acceleration and summation methods are needed most
if only relatively few elements of a slowly convergent or divergent
sequence are available, and if apart from the numerical values of the
input data virtually nothing is known. This is a scenario which happens
only too often if we try to sum divergent perturbation expansions of
quantum physics. In such a situation there is obviously no chance of
constructing analytical expressions for remainder estimates. Instead, we
must construct the remainder estimates from the numerical values of the
input data via simple rules. Fortunately, the simple remainder estimates
proposed by Levin \cite{Levin/1973} and later Smith and Ford
\cite{Smith/Ford/1979}, which are discussed in Section
\ref{Sec:LevExpRemEst}, normally do the job. In spite of their
simplicity, they often work remarkably well.

If we approximate the remainder $r_n$ of a sequence element $s_n$ by the
product $\omega_n z_n$ according to (\ref{Mod_Seq_Om}), where $z_n$ is a
so-called correction term, then we should take into account that our
approximation scheme actually has two degrees of freedom. Levin
\cite{Levin/1973} originally assumed that $z_n$ is a truncated inverse
power series according to (\ref{LevTrModSeq}). This is certainly a very
natural idea, and it leads to a very powerful sequence transformation.

Nevertheless, in some cases Levin's transformation fails horribly for
reasons which we do not completely understand. For example, it was found
in \cite{Weniger/Cizek/Vinette/1991,Weniger/Cizek/Vinette/1993} that
Levin's transformation diverges if it is used for the summation of the
perturbation expansions for the ground state energies of the anharmonic
oscillators (compare also \cite[Table 2]{Weniger/1992} or the discussion
in \cite[Section 10.7]{Weniger/1994b}). A similar divergence of Levin's
transformation was observed by \v{C}\'{\i}\v{z}ek, Zamastil, and
Sk\'{a}la \cite[p.\ 965]{Cizek/Zamastil/Skala/2003} in the case of the
hydrogen atom in an external magnetic field. Fortunately, Levin's choice
for $z_n$ is not the only possibility, and at least for some problems,
alternative correction terms produce significantly better results. For
example, the so-called delta transformation defined in (\ref{dWenTr}) is
based on the assumption that $z_n$ is a truncated factorial series
according to (\ref{WenTrModSeq}). As mentioned before, this delta
transformation is a very effective transformation for slowly convergent
and divergent alternating series. In particular, it produces very good
summation results both in the case of the anharmonic oscillators
\cite{Weniger/1990,Weniger/1992,Weniger/1994b,Weniger/1996c,%
Weniger/1996e,Weniger/Cizek/Vinette/1991,Weniger/Cizek/Vinette/1993} as
well as in the case of the hydrogen atom in an external magnetic field
\cite[Tables I and II]{Cizek/Zamastil/Skala/2003}.

The sequence transformation $\mathcal{G}_{k}^{(n)} (q_m, s_n,
\omega_n)$ introduced by \v{C}\'{\i}\v{z}ek, Zamastil, and 
Sk\'{a}la \cite{Cizek/Zamastil/Skala/2003} permits a unified treatment
of the mathematical properties of all sequence transformations, whose
correction terms are annihilated by difference operators of the type of
\begin{equation}
\hat{T} \; = \; \Delta^k \, P_{k-1} (n) \, .
% \label{}
\end{equation}
Here, $P_{k-1} (n)$ is a suitable polynomial of degree $k-1$ in $n$ that
can be obtained by specializing the parameters $q_m$ in
$\prod_{m=1}^{k-1} (n+q_m)$. All Levin-type transformations considered
in this article belong to this class of sequence
transformations. Consequently, all their mathematical properties such as
explicit expressions (Sections \ref{Sec:AnnihilOp} and
\ref{Sec:RichTypeTrans}), recurrence formulas (Section 
\ref{Sec:RecForm}), and accuracy-through-order properties (Section 
\ref{Sec:RatApprox}) can be deduced from the corresponding properties of
$\mathcal{G}_{k}^{(n)} (q_m, s_n, \omega_n)$ by specializing the 
parameters $q_m$.   

In addition, new sequence transformations can be constructed by
specializing the parameters $q_m$ in $\mathcal{G}_{k}^{(n)} ( q_m, s_n,
\omega_n)$. For example, \v{C}\'{\i}\v{z}ek, Zamastil, and Sk\'{a}la 
\cite[Tables I and II]{Cizek/Zamastil/Skala/2003} found
that in the case of the hydrogen atom in an external magnetic field at
least for some coupling constants better summation results can be
obtained by choosing $q_m = m^2$ instead of choosing $q_m = m$ which
yields the delta transformation with $\beta = 1$. Such a quadratic
dependence of the parameters $q_m$ on $m$ leads to a completely new
sequence transformation. Thus, the sequence transformation
$\mathcal{G}_{k}^{(n)} (q_m, s_n, \omega_n)$ introduced by
\v{C}\'{\i}\v{z}ek, Zamastil, and Sk\'{a}la
\cite{Cizek/Zamastil/Skala/2003} does not only permit a unification of
already known transformations, but it also opens up the path for
promising new research.

As discussed in more details in the following article
\cite{Weniger/2003b*}, our current level of theoretical
understanding does not permit to predict which one of the numerous
variants of $\mathcal{G}_{k}^{(n)} (q_m, s_n, \omega_n)$ will give best
results for a given convergence acceleration or summation problem. So,
if we for example use one of the numerous Levin-type transformation for
the summation of a divergent perturbation expansion, we are essentially
conducting a numerical experiment. As every good experimentalist knows,
a single experiment is only rarely able to provide a definite
answer. Normally, a whole set of related experiments is needed to obtain
convincing evidence. Of course, this applies also to our numerical
experiments. Therefore, we should not insist with a quasi-religious zeal
on using only a single (Levin-type) transformation which we for some
reason may prefer. Instead, it is usually a much better idea to compare
the performance of several different transformations.

Levin-type transformations are not only very powerful but also very
flexible. Experience shows that they can handle successfully a large
variety of different convergence acceleration or summation
problems. This is a direct consequence of the fact that the ansatz
$\omega_n z_n$ for $r_n$ according to (\ref{Mod_Seq_Om}) has two
degrees of freedom which allows a considerable amount of
fine-tuning. Nevertheless, Levin-type transformations are not a cure for
all evils. Consequently, a good experimentalist should also take into
account the possibility that Levin-type transformations may not work at
all or that other transformations could produce better results.

%
% References
%

\newcommand{\SortNoop}[1]{} \newcommand{\OneLetter}[1]{#1}
  \newcommand{\SwapArgs}[2]{#2#1}

\end{document}